\documentclass[twocolumn]{jpsj2}
\usepackage{graphicx}
\usepackage{amsmath,amssymb}

\def\runauthor{Youichi {\sc Yanase}}

\title{ 
Pseudogap and Superconducting Fluctuation 
in High-$T_{\rm c}$ Cuprates: \\
Theory beyond $1$-loop Approximation 
} 

\author{
Youichi {\sc Yanase}\thanks{E-mail: yanase@hosi.phys.s.u-tokyo.ac.jp}
}

\inst{Department of Physics, University of Tokyo, Tokyo 113-0033}

\recdate{December 5, 2003}

\abst
{
 The pseudogap phenomena induced by the SC fluctuation are investigated 
in details. 
 We perform a calculation beyond the $1$-loop approximation. 
 The SC fluctuation is microscopically derived on the basis of 
the repulsive Hubbard model. 
 The vertex corrections are collected in the infinite order 
with use of the quasi-static approximation. 
 The single-particle excitations, NMR $1/T_{1}T$, spin susceptibility 
and superconducting transition temperature are discussed. 
 The important role of the vertex correction is pointed out for the 
single particle spectral function. On the other hand, 
the validity of the $1$-loop order theory is confirmed for 
other quantities.  
 We shed light on the essential nature of SC fluctuation leading 
to the pseudogap from the comparison with spin and charge fluctuations. 
}

\kword
{Superconducting fluctuation; spin fluctuation; 
vertex correction; high-$T_{\rm c}$ cuprates; pseudogap
}

\begin{document}
\sloppy
\maketitle

\newcommand{\nozi}{Nozi$\acute{{\rm e}}$res }
\newcommand{\eli}{$\acute{{\rm E}}$liashberg }
\renewcommand{\k}{\mbox{\boldmath$k$}}
\newcommand{\p}{\mbox{\boldmath$p$}}
\newcommand{\q}{\mbox{\boldmath$q$}}
\newcommand{\Q}{\mbox{\boldmath$Q$}}
\newcommand{\kk}{\mbox{\boldmath$k'$}}
\newcommand{\e}{\varepsilon}
\newcommand{\ee}{\varepsilon^{'}}
\newcommand{\s}{{\mit{\it \Sigma}}}
\newcommand{\J}{\mbox{\boldmath$J$}}
\newcommand{\vv}{\mbox{\boldmath$v$}}
\newcommand{\Jh}{J_{{\rm H}}}
\newcommand{\LL}{\mbox{\boldmath$L$}}
\renewcommand{\SS}{\mbox{\boldmath$S$}}
\newcommand{\Tc}{$T_{\rm c}$ }
\newcommand{\Tcc}{$T_{\rm c}$}
\newcommand{\Tmf}{$T_{\rm c}^{\rm MF}$ }
\newcommand{\bmk}{\mbox{\scriptsize \boldmath$k$}}
\newcommand{\bmkk}{\mbox{\scriptsize \boldmath$k'$}}
\newcommand{\bmq}{\mbox{\scriptsize \boldmath$q$}}

\section{Introduction}

\begin{figure}[t]
\begin{center}
\includegraphics[height=5cm]{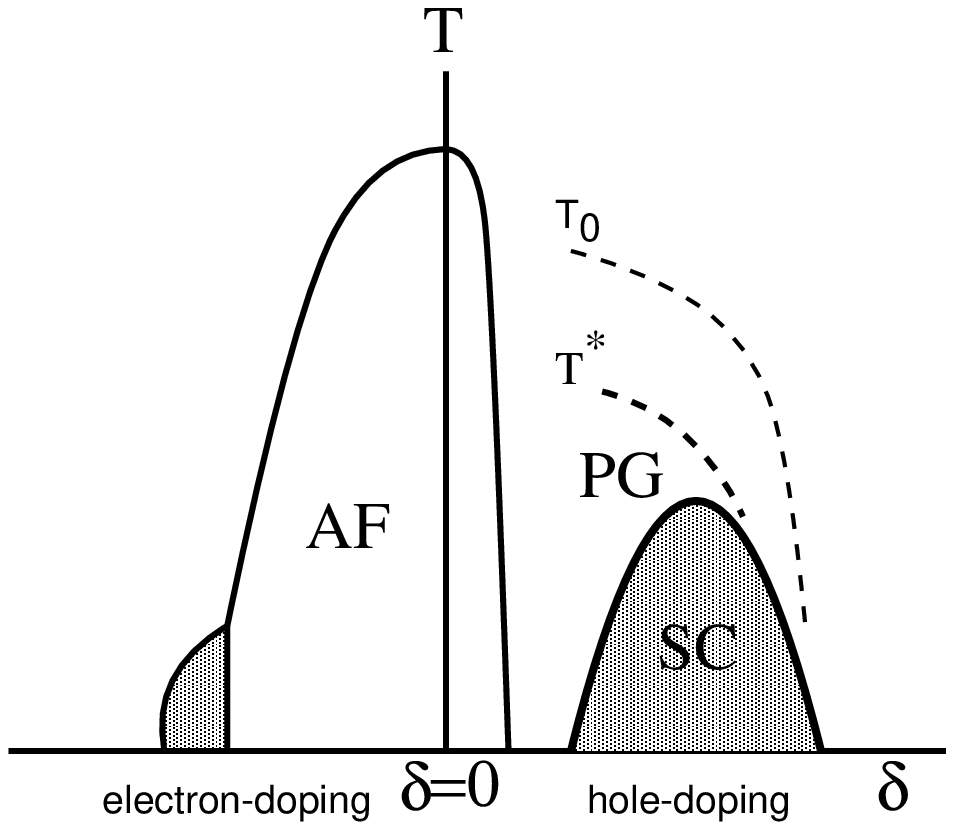}
\caption{The phase diagram of high-$T_{\rm c}$ superconductors.
The horizontal and vertical axes indicate the doping concentration
and the temperature, respectively.
``AF'', ``SC'', and ``PG'' denote anti-ferromagnetic, superconducting 
and pseudogap state, respectively.
The onset curve for the spin fluctuation ($T$=$T_0$) and that for
the SC fluctuation ($T$=$T^*$) show cross-over 
temperatures.} 
\label{fig:high-Tcphasediagram2}
\end{center}
\end{figure}

 Since the discovery in 1986~\cite{rf:bednortz}, many anomalous aspects 
of high-$T_{\rm c}$ superconductors have been clarified by 
enormous studies. 
 Among them, the pseudogap phenomena~\cite{rf:timusk} 
(Fig.~\ref{fig:high-Tcphasediagram2}) have indicated a break down of 
the Fermi liquid theory, which is a universal view on the low energy 
excitations in metals~\cite{rf:nozieresbook}. 
 The unusual nature has indicated 
an appearance of a new concept in the condensed matter 
physics~\cite{rf:andersonbook}.

 From the experimental point of view, the pseudogap was found 
in the magnetic excitation by the nuclear magnetic resonance 
(NMR)~\cite{rf:yasuoka}. 
 At present anomalies have been observed in many experimental 
probes~\cite{rf:timusk} which include 
NMR~\cite{rf:yasuoka,rf:warren,rf:takigawa1991,rf:itoh1992,rf:itoh1998,
rf:ishida1998,rf:alloul}, 
neutron scattering~\cite{rf:neutronPG}, 
electric transport~\cite{rf:iyereview,rf:ito,rf:mizuhashi,
rf:odatransport,rf:satoM,rf:takenaka}, 
optical spectrum~\cite{rf:homes,rf:basov,rf:tajima}, 
tunneling spectroscopy~\cite{rf:renner,rf:miyakawa,rf:dipasupil}  
and angle resolved photo emission spectroscopy 
(ARPES)~\cite{rf:ding,rf:shenPG,rf:normanPG,rf:ARPESreview}.

 Among many theoretical 
proposals~\cite{rf:baskaranRVB,rf:fukuyamaRVBreview,rf:nagaosa,
rf:kampfPG,rf:chubukovPG,rf:vilkAFPG,rf:chakravarty}, 
we have adopted 
the ``pairing scenario'' in which the pseudogap is 
induced by the superconducting (SC) fluctuation. 
 This scenario has been indicated by several experimental results 
cited above. 
 In particular, ARPES, which presents a momentum 
resolved information, has provided a strong circumstantial evidence. 
 The pseudogap in the single particle excitation first opens around 
the zone boundary $\k \sim$ ($\pi,0$) and gradually extends 
to the whole Fermi surface as approaching to the 
superconducting transition. 
 The amplitude of the excitation gap does not change through \Tcc. 
 The latter has been observed also in the tunneling 
spectroscopy~\cite{rf:renner,rf:miyakawa,rf:dipasupil} and 
in the $c$-axis optical conductivity~\cite{rf:homes,rf:basov,rf:tajima}. 
 These results have indicated a close relation between the 
superconductivity and pseudogap.

 Since there are many approaches included in the pairing scenario, 
we have classified them into three kinds in order to avoid any 
confusion.~\cite{rf:yanasereview}  
 One is the BCS-BEC cross-over 
~\cite{rf:micnasreview,rf:melo,rf:randeriareview,rf:haussmann}
formulated in the \nozi and Schmitt-Rink theory~\cite{rf:Nozieres}. 
 However, this approach is not relevant for high-$T_{\rm c}$ cuprates 
because this is basically justified in the low density system. 
 We have adopted the second approach which is the perturbation theory 
with respect to the SC fluctuation.~\cite{rf:janko,rf:yanasePG,
rf:metzner,rf:perali3D,rf:yanasereview}  
 Then, the resonance scattering described by the electron self-energy 
plays an essential role. 
 The $1$-loop order theory which is called T-matrix approximation 
has been widely used and enabled us to perform a non-phenomenological 
treatment on the microscopic Hamiltonian. 
 This approach is complementary with the third one, 
phase fluctuation theory.~\cite{rf:emery,rf:franz,rf:kwon,rf:phase,
rf:eckl,rf:tesanovic} 
 The amplitude of order parameter is fixed 
in the phase fluctuation theory, which will be justified in 
the non-perturbative region with respect to the higher order 
mode coupling.

 It is our basic standpoint that non-Fermi liquid behaviors 
in cuprates should be derived from the Fermi liquid state. 
 Because cuprates behave as a conventional metal 
in the over-doped region and there is no discontinuity between
over-doped and under-doped regions, the origin of the anomalous 
behaviors should be inherent in the Fermi liquid state. 
 Such understanding will be valuable for a universal understanding of 
the strongly correlated electron systems including transition metals, 
organic superconductors and heavy fermion systems. 
 Then, the Fermi liquid theory will be a reasonable starting point. 
 The microscopic study is suitable for this purpose 
since the characteristics of each system should be clarified from the 
microscopic point of view.

 In the previous studies,~\cite{rf:yanaseFLEXPG,rf:yanaseTRPG} 
a microscopic theory for the SC fluctuation is developed on the 
basis of the repulsive Hubbard model. Then, the T-matrix and 
self-consistent T-matrix approximations for the SC fluctuation 
have been used in combination with the FLEX approximation~\cite{rf:FLEX}. 
 It has been shown that single-particle, magnetic and 
transport properties are explained in a coherent way. 
 The doping dependence including the particle-hole asymmetry is also 
reproduced without using any phenomenological assumption. 
 The electron-doped cuprates are basically weak-coupling superconductors 
and do not show pseudogap. 
 The pseudogap state in the hole-doped region is described as 
a regime with spin and SC fluctuations, which have both 
cooperative and competitive aspects. 
 We have shown that the pseudogap is essentially induced by 
the SC fluctuation.

 The purpose of this study is an examination of the theory based on 
the $1$-loop approximation. 
 The previous studies have attributed the pseudogap to the serious 
correction arising from the $1$-loop order term.~\cite{rf:janko,
rf:yanasePG,rf:metzner,rf:perali3D,rf:yanasereview,rf:yanaseFLEXPG,
rf:yanaseTRPG}  
 This fact involves a naive question from the theoretical point of view. 
 Do the higher order terms play a more serious role ? 
 There is no clear evidence for the validity of $1$-loop 
order theory because so-called Migdal theorem is not generally applicable. 
 In this paper we present a theory beyond the $1$-loop approximation 
in order to answer this question. 
 This study will clarify the roles of higher order corrections.

 Theoretical studies beyond the $1$-loop approximation have been 
developed mainly for the Lee-Rice-Anderson model~\cite{rf:leericeanderson} 
which is an one-dimensional model with charge fluctuation. 
 Then, the electronic system coupled to the classical Gaussian 
field has been investigated.~\cite{rf:sadovskii,rf:tchernyshyov1D,
rf:kopietzPG,rf:millisPG,rf:monienPG}  
 The exact estimation is possible in the limit of 
infinite correlation length.~\cite{rf:sadovskii}  
 An approximation for the finite correlation
length has been suggested by Sadovskii~\cite{rf:sadovskii} and 
its validity has been investigated extensively.~\cite{rf:tchernyshyov1D, 
rf:kopietzPG,rf:millisPG,rf:monienPG} 
 The theoretical method developed there has been applied to 
the two dimensional systems with charge fluctuation,~\cite{rf:kuchinskii} 
spin fluctuation~\cite{rf:schmalianPG} and 
SC fluctuation,~\cite{rf:tchernyshyov} respectively. 
 We will explain in \S3 that this method is formulated in 
the microscopic theory for the Hubbard-type Hamiltonian.  
 It will be discussed that the estimation based on the infinite 
correlation length is applicable more robustly in the present case. 
 We apply this method as well as the Sadovskii's method 
to the repulsive Hubbard model in \S4. 
 The attractive Hubbard model is also briefly adopted as another typical 
example in \S5.1.

 We will show that higher order corrections are renormalized to be small, 
even if higher order terms develop in the scheme of 
naive perturbation calculation. 
 It is concluded that the $1$-loop order theory is 
qualitatively valid in many aspects (\S4.2-4). 
 On the other hand, the importance of the vertex correction 
will be illuminated for the single-particle spectral function, 
for which the $1$-loop and self-consistent $1$-loop 
approximations provide qualitatively incompatible results. 
 It is shown that the non-self-consistent $1$-loop approximation 
is qualitatively correct for this quantity (\S4.1 and \S5.1). 
 This is an example in which the partial summation included in 
the self-consistent theory provides an unphysical result. 
 We will show that this conclusion is non-trivial 
by comparing with the case of spin fluctuation and that of 
charge fluctuation (\S5.2). 
 Indeed, the roles of vertex corrections are determined 
by the symmetry of the order parameter. 
 We furthermore illustrate that the differences between the 
order parameters are qualitatively understood from the naive perturbation 
theory (\S5.3).

 This paper is constructed in the following way. 
 The $1$-loop order theory is shortly summarized in \S2.  
 Higher order terms are classified in \S3. 
 We introduce the terms focused in this paper and formulate 
the infinite-order theory. 
 A justification for the adopted approximation is given in Appendix A and B. 
 The roles of vertex corrections are investigated in \S4. 
The single particle excitations (\S4.1), superconducting \Tc (\S4.2-3)
and magnetic properties (\S4.4) are discussed in details. 
 The vertex corrections beyond the formulation in \S3 are briefly 
discussed in the estimation of \Tc (\S4.3). 
 The Sadovskii's method is used in \S4.4, where the naive perturbation 
is performed within the $4$-loop order 
and compared with the Sadovskii's method for a justification. 
 The case of charge fluctuation and that of spin fluctuation are 
over-viewed in \S5, where the essential nature of SC fluctuation is 
clarified. 
 The summary and discussion are given in the last section \S6.

\section{Theory in the $1$-loop order}

 In this section, the  $1$-loop order theory is 
summarized before discussing the vertex corrections. 
 After the basic mechanism of the pseudogap was described 
on the basis of the attractive model,~\cite{rf:janko,rf:yanasePG,
rf:metzner,rf:perali3D} the microscopic theory has been developed 
on the basis of the repulsive Hubbard model~\cite{rf:yanaseFLEXPG,
rf:yanaseTRPG,rf:dahmPG}. 
 An approach for the infinite-$U$ $d$-$p$ model has been also 
developed.~\cite{rf:kobayashi} 
 Here we explain the formulation and typical results of the repulsive 
Hubbard model for a discussion in the following 
sections.~\cite{rf:commentattractive}

 The Hubbard Hamiltonian is expressed as,  
\begin{eqnarray}
  \label{eq:Hubbard-model}
   H=\sum_{{\k},\sigma} \varepsilon(\k) 
  c_{{\k},\sigma}^{\dag}c_{{\k},\sigma}
  + U \sum_{i} n_{{i},\uparrow} n_{{i},\downarrow}. 
\end{eqnarray}
 We consider the square lattice and choose the following tight-binding 
dispersion $\varepsilon(\k)$, 
\begin{eqnarray}
  \label{eq:high-tc-dispersion}
   \varepsilon(\k)=-2t(\cos k_{\rm x}+\cos k_{\rm y})
  +4t'\cos k_{\rm x} \cos k_{\rm y}-\mu. 
\end{eqnarray}
 The parameters are chosen as $2t=1$, $t'/t=0.25$, $U/t=4.2$, 
and the doping concentration is chosen to be $\delta=1-n=0.1$,  
unless we mention explicitly.

\begin{figure}[t]
\begin{center}
\includegraphics[height=4.5cm]{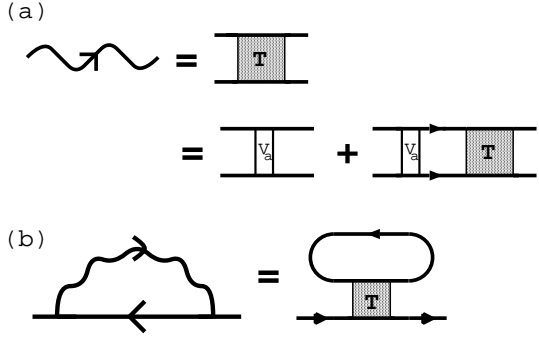}
\caption{(a) The diagrammatic representation of the T-matrix. 
We denote the effective interaction between quasi-particles as $V_{\rm a}$. 
(b) The self-energy in the T-matrix approximation. 
} 
\label{fig:T-matrix}
\end{center}
\end{figure}

 The SC fluctuation is generally described by the T-matrix 
which is the propagator of SC fluctuation. 
 The T-matrix is represented by the ladder diagram as 
Fig.~\ref{fig:T-matrix}(a) where the effective interaction 
in the Cooper channel is represented by $V_{\rm a}$. 
 In the attractive models, the first order perturbation has been used for 
the estimation of $V_{\rm a}$. ~\cite{rf:janko,rf:yanasePG,
rf:metzner,rf:perali3D}
 In order to describe the superconductivity in the repulsive Hubbard model, 
we have to derive the effective interaction from the many body effects. 
 Here we adopt the FLEX approximation~\cite{rf:FLEX} which has been 
widely used as a microscopic description of the nearly 
anti-ferromagnetic Fermi liquid.~\cite{rf:moriyaAD,rf:chubukovreview} 
 The following calculation can be regarded as an improvement of the 
FLEX approximation so as to take account of the SC fluctuation. 
 Then, the effective interaction is described by the spin and charge 
susceptibility in the following way, 
\begin{eqnarray}
  \label{eq:anomalous-rpa-singlet}
  V_{\rm a}(k,k')=U + \frac{3}{2} U^{2} \chi_{{\rm s}}(k-k') 
  -\frac{1}{2} U^{2} \chi_{{\rm c}}(k-k'),
\end{eqnarray}
where spin and charge susceptibility is estimated as, 
\begin{eqnarray}
  \label{eq:rpa-susceptibility}
  && \hspace{-10mm} \chi_{{\rm s}}(q) = \frac{\chi_{0}(q)}{1 - U \chi_{0}(q)},
  \hspace{5mm}
  \chi_{{\rm c}}(q) = \frac{\chi_{0}(q)}{1 + U \chi_{0}(q)}, 
\\
  \label{eq:irreducible-susceptibility}
  &&  \hspace{-10mm} \chi_{0}(q) = -\sum_{k} G(k+q) G(k). 
\end{eqnarray} 
 Here, the renormalized Green function is expressed as,  
\begin{eqnarray}
  \label{eq:dressed-Green-function}
  G(k)=\frac{1}{{\rm i} \omega_{n} - \varepsilon(\k) - \Sigma (k)}. 
\end{eqnarray}
 The self-energy in the FLEX approximation is obtained as 
$\Sigma (k)=\Sigma_{\rm F} (k)$, where  
\begin{eqnarray}
  \label{eq:rpa-normal}
  \Sigma_{\rm F} (k) = \sum_{q} V_{\rm n} (q) G(k-q),
\\
  \label{eq:rpa-normal-vertex}
   V_{\rm n} (q)=U^{2} [\frac{3}{2} \chi_{{\rm s}} (q)
  +\frac{1}{2} \chi_{{\rm c}} (q)-\chi_{0} (q)].
\end{eqnarray}
 With use of these functions, the superconducting \Tc is estimated by 
the following eigenvalue equation (linearized \eli equation), 
\begin{eqnarray}
  \label{eq:eliashberg-super}
   \lambda_{{\rm e}} \phi(k) =
  -\sum_{k'} V_{{\rm a}}(k,k') |G(k')|^{2} \phi(k'). 
\end{eqnarray}
 The transition temperature is determined by the criterion that 
the maximum eigenvalue is unity ($\lambda_{\rm e}$=1) at $T=T_{\rm c}$. 
 The FLEX approximation is regarded to be the mean field theory 
in the sense that the SC fluctuation is not taken into account. 
 It should be stressed that the \eli equation is derived from the  
strong coupling theory for the superconductivity which is needed 
for the quantitative estimation of \Tcc. 
 This effort is important for the present purpose because the value of 
\Tc plays an essential role for the SC fluctuation.~\cite{rf:yanasereview} 
 Note that the weak coupling theory significantly over-estimates the value 
of \Tcc.

 The T-matrix has been estimated by using the extended \eli equation as, 
 \begin{eqnarray}
  && \hspace{-10mm} T(k_{1},k_{2}:q)=\phi(k_{1}) t(q) \phi^{*}(k_{2}),  
  \label{eq:FLEX+T-matrix1} 
\\ 
  && \hspace{-10mm}  t(q)=\frac{g \lambda(q)}{1-\lambda(q)}, 
\\
  && \hspace{-10mm} \lambda(q)=
  -\sum_{k,p}\phi^{*}(k) V_{\rm a}(k-p) G(p) G(q-p) \phi(p), 
  \label{eq:FLEX+T-matrix2}
\\
&& \hspace{-10mm}  
g=\sum_{k_{1},k_{2}}\phi^{*}(k_{1}) V_{\rm a}(k_{1}-k_{2}) \phi(k_{2}),
\end{eqnarray}
where the wave function is normalized as 
\begin{eqnarray}
\label{eq:normalization}
  \sum_{k} |\phi(k)|^{2} = 1.
\end{eqnarray}
 Since this estimation of the T-matrix is correct around 
$q=0$, the role of SC fluctuation is appropriately 
represented.~\cite{rf:yanaseFLEXPG} 
 It should be noted that the parameter $\lambda(0)$ is equivalent 
to the maximum eigenvalue of the \eli equation, 
namely $\lambda(0)=\lambda_{{\rm e}}$. 
 Thus, the divergence of the T-matrix is equivalent to the criterion 
in the \eli equation.

 The self-energy correction due to the SC fluctuation 
is obtained within the $1$-loop order as (Fig.~\ref{fig:T-matrix}(b)), 
\begin{eqnarray} 
\label{eq:FLEX+T-matrixap}
 \Sigma_{{\rm S}}(k)=\sum_{q} T(k,k:q) G(q-k).
\end{eqnarray} 
 The total self-energy is obtained by adding it to the contribution
from the spin fluctuation,  
\begin{eqnarray} 
\label{eq:total-self-energy}
\Sigma(k)=\Sigma_{\rm F}(k)+\Sigma_{\rm S}(k). 
\end{eqnarray} 
 The lowest order theory has been denoted as FLEX+T-matrix approximation 
(FTA) while the self-consistent calculation has been denoted as 
self-consistent FLEX+T-matrix approximation (SCFT). 
 The basic role of the SC fluctuation is described by the FTA which is 
the $1$-loop approximation with respect to the SC fluctuation. 
 The SCFT is the self-consistent $1$-loop order theory 
and furthermore includes the feedback effect on the magnetic 
fluctuation. 
 The SCFT is needed to explain the transport phenomena 
in which the feedback effect plays an essential role~\cite{rf:yanaseTRPG}.

\begin{figure}[htbp]
  \begin{center}
\includegraphics[height=10cm]{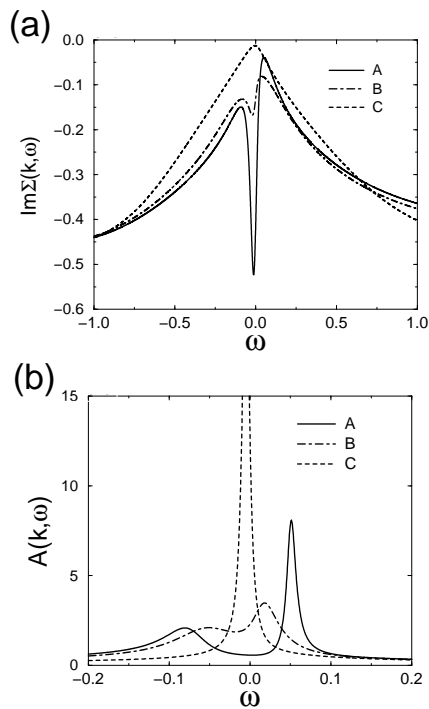}
    \caption{(a) The imaginary part of the self-energy 
             and (b) the spectral function on the Fermi surface. 
             A, B and C show the results at $\k=(0.07\pi,\pi)$, 
             $\k=(0.15\pi,0.8\pi)$ and $\k=(0.43\pi,0.45\pi)$, respectively. 
             The parameters are chosen as $U/t=3.2$, $\delta=0.1$ and 
             $T=1.2 T_{{\rm c}}$. 
             }
    \label{fig:1-loop}
  \end{center}
\end{figure}

 Figure.~\ref{fig:1-loop} shows the spectral function and 
the self-energy obtained by FTA. 
 We clearly see the anomalous behaviors. 
 The imaginary part of the self-energy (Fig.~\ref{fig:1-loop}(a)) 
shows the scattering peak of 
$|{\rm Im} {\mit{\it \Sigma}}^{{\rm R}} (\mbox{\boldmath$k$},\omega)|$ 
around $\omega=0$. 
 This feature is qualitatively different from the conventional Fermi 
liquid where the self-energy behaves as 
$|{\rm Im} {\mit{\it \Sigma}}^{{\rm R}} (\mbox{\boldmath$k$},\omega)|
\propto \omega^{2}$. 
 In this sense, the Fermi liquid theory breaks down 
in the pseudogap state owing to the SC fluctuation. 
 The scattering peak around $\omega=0$ leads to the pseudogap 
around \k$=(\pi,0)$ as is shown in Fig.~\ref{fig:1-loop}(b). 
 This is the basic mechanism of the pseudogap phenomena induced by the 
SC fluctuation. 
  The pseudogap gradually occurs with decreasing the temperature 
and/or doping concentration.

 Concerning the momentum dependence, the Fermi liquid behaviors 
appear around \k$=(\pi/2,\pi/2)$ which is denoted as ``cold spot''  
in contrast to the ``hot spot'' around \k$=(\pi,0)$. 
 The $\omega$-linear behavior instead of $\omega$-square behavior 
is induced by the spin fluctuation. 
 This momentum dependence induces the ``Fermi arc'' which has been 
observed in ARPES.~\cite{rf:ding,rf:shenPG,rf:normanPG,rf:ARPESreview} 
 Note that these anomalous behaviors are suppressed in the electron-doped 
cuprates where the SC fluctuation is very weak.~\cite{rf:yanaseFLEXPG,
rf:yanasereview}

 Because the SC fluctuation is characterized by 
the T-matrix around $\mbox{\boldmath$q$}=\Omega=0$, 
the following expansion (TDGL expansion) is instructive for a 
qualitative understanding.
\begin{eqnarray}
  \label{eq:TDGL}
  t(\mbox{\boldmath$q$},\Omega)=
  \frac{g}{t_{0}+b\mbox{\boldmath$q$}^{2}-(a_{1}+{\rm i}a_{2})\Omega}.
\end{eqnarray}
 The parameter $ b $ is generally related to the coherence length 
$\xi_{0}$ as $b \propto \xi_{0}^{2}$, which is estimated as 
$\xi_{0} \propto v_{{\rm F}}/T_{\rm c}^{\rm MF}$ in the clean limit.  
 Here $T_{\rm c}^{\rm MF}$ is the transition temperature in the mean 
field theory. 
 We understand that sufficiently large value of $b$ is obtained in the 
weak-coupling superconductor. 
 Then, the total weight of the SC fluctuation is generally small. 
 This is the reason why the fluctuation is negligible in conventional
superconductors. 
 In this sense, the over-doped and electron-doped cuprates are 
classified into the weak-coupling superconductor. 
 We have obtained the sufficiently large value of $b$ in these 
regions.~\cite{rf:yanaseFLEXPG} 
 On the contrary, the coherence length in the under-doped region 
has been estimated as $\xi_{0} = 3 \sim 5$ in the unit of the lattice 
constant. 
 This is the reason why the SC fluctuation plays an important role.

 At the last of this section, we comment on the difference between 
FTA and SCFT. 
In the SCFT the dressed Green function is assigned in 
eqs.~\ref{eq:irreducible-susceptibility}-\ref{eq:FLEX+T-matrixap} 
instead of the Green function obtained by the FLEX approximation. 
 Then, the pseudogap similarly appears in most of quantities, 
such as DOS,~\cite{rf:yanaseFLEXPG} 
although effects of the SC fluctuation are reduced. 
 However, the situation is quite different for the single particle 
spectral function, which has been widely interested 
in the studies of high-\Tc cuprates.  
 The SCFT does not show the pseudogap in this quantity on 
the Fermi surface, while it does slightly apart from the Fermi surface. 
 This discrepancy will be discussed in \S4.1 and \S5.1. 
 This issue is particularly important because the pseudogap 
in the spectral function is a novel characteristics of the strong 
coupling superconductor as well as because it is an experimental 
evidence~\cite{rf:ding,rf:shenPG,rf:normanPG,rf:ARPESreview} 
for the pairing scenario.

\section{Vertex correction}

 In this section we classify the vertex corrections 
and formulate the infinite order theory. 
 Most part of this section is provided from the general point of view. 
 The validity of approximations adopted here will be examined on the 
basis of the microscopic theory explained in \S2 and developed in \S4, 
which is relevant for high-\Tc cuprates.

\subsection{Classification of vertex corrections}

\begin{figure}[htbp]
  \begin{center}
\includegraphics[height=2.7cm]{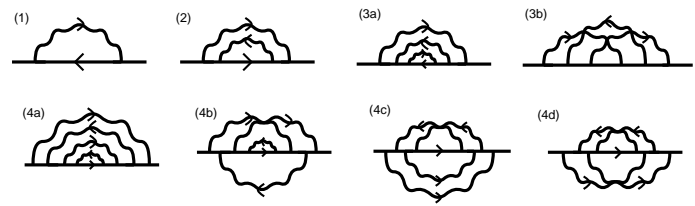}
    \caption{The series of vertex corrections focused 
            in this paper. 
            The solid and wavy lines represent the propagator of  
            electron and SC fluctuation, respectively. 
             }
\label{fig:yokohigh}
  \end{center}
\end{figure}

 We adopt a perturbation expansion with respect to the coupling 
between electrons and SC fluctuation, which is called loop expansion. 
 In more general perturbation scheme for the Hubbard 
Hamiltonian,~\cite{rf:AGD} there appear other terms which are not 
included in the loop expansion. 
 However, we simply ignore them and focus on 
the singular terms in the vicinity of the phase transition. 
 The vertex corrections to the Green function are mainly discussed in 
this subsection. 
 In addition to them, the vertex corrections to the pairing correlation 
function will be investigated in \S4.3.

\begin{figure}[htbp]
  \begin{center}
\includegraphics[height=4cm]{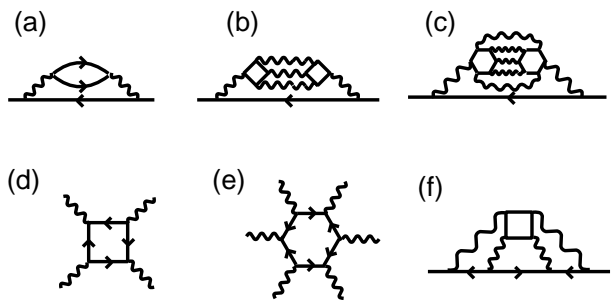}
    \caption{Examples of diagrams which are not included 
             in Fig.~\ref{fig:yokohigh}. 
             (a) A diagram included in the $1$-loop order term 
                 owing to the definition. 
             (b), (c) and (f) are diagrams including higher order 
                 mode couplings. 
             (d) and (e) are the forth order and sixth order vertex, 
                 respectively. 
             }
\label{fig:highervertex}
  \end{center}
\end{figure}

 First of all, we show the Feynmann diagrams corresponding to the 
self-energy investigated in this paper (Fig.~\ref{fig:yokohigh}). 
 Fig.~\ref{fig:yokohigh}(1) is equivalent to Fig.~\ref{fig:T-matrix}(b), 
and Figs.~\ref{fig:yokohigh}(2), (3a) and (4a) are included in 
the self-consistent T-matrix approximation. 
 The other terms are classified into the vertex correction. 
 The diagrams including the electron loop are not included in 
Fig.~\ref{fig:yokohigh}. 
 It should be noticed that the terms like Fig.~\ref{fig:highervertex}(a) 
should be included in the $1$-loop order term owing to the definition. 
 Considering this fact, we understand that all of the ignored terms 
include the higher order mode couplings between SC fluctuations,  
which correspond to the non-linear terms in the Ginzburg Landau expansion 
($4$th, $6$th, $8$th..... order terms). 
 For example, Fig.~\ref{fig:highervertex}(b) and 
Fig.~\ref{fig:highervertex}(c) include 
the fourth order vertex (Fig.~\ref{fig:highervertex}(d)) and 
sixth order vertex (Fig.~\ref{fig:highervertex}(e)), respectively. 
 Therefore, the estimation of the terms in Fig.~\ref{fig:yokohigh} is 
sufficient in the renormalized Gaussian region where the higher order 
mode coupling does not play any essential role, namely when the system is 
not too close to the critical point. 
 We note that the mode coupling effect in the level of the 
self-consistent renormalization theory (SCR)~\cite{rf:moriyaAD} is 
included by itself if we determine the fluctuation propagator 
self-consistently as in \S4.2-4.

 The diagrammatic expression in Fig.~\ref{fig:yokohigh} is exact 
if we consider the Lee-Rice-Anderson model. 
 Most of the previous studies beyond the $1$-loop 
approximation~\cite{rf:sadovskii,rf:tchernyshyov1D,
rf:kopietzPG,rf:millisPG,rf:monienPG,rf:kuchinskii,rf:schmalianPG,
rf:tchernyshyov} have been devoted to this 
phenomenological model. 
 The importance of these terms has been pointed out for 
the model with sufficiently long range attractive interaction 
where the calculation is reduced to 
the zero-dimensional problem.~\cite{rf:fujimotoPG} 
 We will estimate these terms as an approximation 
for the microscopic model.

 The higher order mode couplings are surely important   
when the system is close to the critical point. 
 However, considerable part of them is represented by the 
renormalization of the fluctuation propagator. 
 For example, the terms in Figs.~\ref{fig:highervertex}(b) and (c) 
are the cases. 
 These corrections are represented by the TDGL parameters 
(see eq.~\ref{eq:TDGL}) and the shift of \Tcc. 
 Therefore, the qualitative roles of SC fluctuation are not affected, 
although these corrections may be important for the estimation of \Tcc. 
 In this sense we have mentioned that the corrections  
in Fig.~\ref{fig:yokohigh} are important in the ``renormalized'' 
Gaussian region. 
 Indeed, the renormalization of the fluctuation propagator has been 
investigated from 1980's motivated by the theoretical issues on 
the BCS-BEC cross-over.~\cite{rf:tokumitu,rf:sato}  
 Then, the higher order terms arising from the forth order vertex 
(Fig.~\ref{fig:highervertex}(d)) have been investigated and 
concluded to be not important quantitatively. 
 We will investigate the different kinds of mode coupling in order to 
discuss the doping dependence of \Tc in \S4.3. 
 They include the lower order terms with respect to 
the fluctuation propagator, rather than the terms investigated 
in Refs.71 and 72.

 There remain corrections including higher order mode couplings 
but not discussed above. 
 We show the lowest order term in Fig.~\ref{fig:highervertex}(f) 
which is the $4$-loop order term. 
 This term is more singular rather than the same order terms 
in Fig.~\ref{fig:yokohigh}.  
 However, we will estimate this term,  
and conclude that this correction is small 
in the wide region of the pseudogap state (Appendix B).

\subsection{Quasi-static approximation}

 According to these discussions, we consider the vertex 
corrections represented by Fig.~\ref{fig:yokohigh}. 
 We adopt the quasi-static approximation in order to formulate 
an infinite-order theory. 
 In this approximation, the thermal fluctuation is taken into account 
but the quantum aspect of the fluctuation is ignored. 
 This procedure has been adopted in the studies of critical phenomena. 
 In the formulation, the zero Matsubara frequency $\Omega_{n}=0$ of 
the fluctuation propagator is taken into account. 
 This procedure is generally justified at high temperature and/or 
in the vicinity of the critical point. 

 Indeed, the character of SC fluctuation emarges in the applicability of 
the quasi-static approximation. 
 The condition for the quasi-static approximation is generally described as 
$\omega_{\rm sc} \ll T$ where $T$ is the temperature and 
$\omega_{\rm sc}$ is the characteristic frequency of the SC fluctuation. 
 Interestingly, $\omega_{\rm sc}$ is expressed as 
$\omega_{\rm sc} \sim T \varepsilon$, where 
$\varepsilon = (T-T_{\rm c})/T_{\rm c}$ is the reduced temperature. 
 Then, the quasi-static approximation is justified when $\varepsilon \ll 1$. 
 This fact should be contrasted to the case of AF spin fluctuation. 
 Then, the characteristic frequency of the fluctuation is 
expressed as $\omega_{\rm s} \sim E_{\rm F} \varepsilon_{\rm s}$, 
where $\varepsilon_{\rm s}$ 
denotes the reduced temperature for the magnetic instability. 
 It should be noticed that the factor appearing in $\omega_{\rm sc}$ 
is smaller than that in $\omega_{\rm s}$ by the order $T/E_{\rm F}$. 
 This fact implies the better accuracy of quasi-static approximation 
for the SC fluctuation rather than that for the spin fluctuation. 
 This will be confirmed by the microscopic calculation based on the 
repulsive Hubbard model (see Appendix A). 
 While the quasi-static approximation for spin fluctuation 
is basically justified at high temperature, that for 
SC fluctuation is effective in much lower temperature region.

 Another interpretation of the quasi-static approximation is obtained 
from the application to the $1$-loop order theory. 
 In \S2, we have shown the anomalous behavior 
of the self-energy induced by the SC fluctuation. 
 It is easily confirmed that this anomaly arises from 
the quasi-static part of the T-matrix $t(\q,\Omega_{n}=0)$, 
which is singular in the vicinity of \Tcc. 
 The remaining contribution from the dynamical part 
basically induces the Fermi liquid behavior.  
 Thus, anomalous contribution leading to the pseudogap is 
appropriately included in the quasi-static approximation. 
 Indeed, the applicability of quasi-static approximation is 
a requirement of the pseudogap phenomena because the pseudogap in 
the single particle spectral function appears when the former is 
more significant than the latter. 
 Therefore, the quasi-static approximation is sufficient for the purpose 
of this paper, while the closing of pseudogap may be described 
inadequately.

 Owing to the quasi-static approximation, only the momentum summation 
remains as a practical task. 
 Although an exact estimation is still difficult, 
we can avoid this difficulty by considering the ``dirty'' region where 
the quasi-particle mean free path is much smaller than the GL  
correlation length, namely $l \ll \xi_{\rm GL}$. 
 In this region, the $q$-dependence of the Green function can be ignored, 
where $q$ is the momentum of fluctuation. 
 It should be noticed that quasi-particle in the ``hot spot'' has a 
remarkably short mean free path owing to strong scattering 
due to the AF spin fluctuation.~\cite{rf:hlubina,
rf:stojkovic,rf:yanaseTR} 
 Therefore, the condition $l \ll \xi_{\rm GL}$ is relevant in this region. 
 Note that the superconductivity and magnetic properties are 
dominated by this region of the Fermi surface, while 
transport phenomena are not.~\cite{rf:yanaseTRPG}

 In the ``dirty'' region, the Green function in the $n$-loop order 
is equivalently estimated as, 
\begin{eqnarray}
  \label{eq:infinite-series-SC}
G^{(n)}(k) = (-1)^{n} \Delta^{2 n}(k) G^{(0)}(k)^{n+1} G^{(0)}(-k)^{n}, 
\end{eqnarray}
where  $\Delta^{2}(k)= \Delta^{2} |\phi(k)|^{2}$ and 
$\Delta^{2}= T \sum_{q} |t(q,0)|$, which is proportional to 
the thermal weight of the SC fluctuation. 
 The equivalence of $n$-loop order diagrams will be briefly 
discussed in \S4.4 where a justification is obtained from 
the numerical estimation. 
 We have defined the Green function without including the SC fluctuation 
as $G^{(0)}(k)$. 
 In the present case, the renormalization from the spin fluctuation 
is taken into account as $G^{(0)}(k)=G_{\rm F}(k)=
({\rm i}\omega_{n} - \varepsilon(\k)-\Sigma_{\rm F}(k))^{-1}$.

 Counting the topological factor $n!$, the summation of the infinite 
series is obtained as an Eular's error function,~\cite{rf:tchernyshyov} 
which is numerically tractable with use of the integral representation, 
\begin{eqnarray}
  \label{eq:infinite-loop}
 G(k)=G_{\rm F}(k) \int_{0}^{\infty} \frac{e^{-t}}{1+t \Delta^{2}(k) 
|G_{\rm F}(k)|^{2}} {\rm d}t. 
\end{eqnarray} 
 In this procedure the properties of the SC fluctuation are 
represented by the single parameter $\Delta$. 
 We use this approximation in \S4.1-3 where the single particle 
properties around $T=T_{\rm c}$ and the value of \Tc are investigated. 
 A more sophisticated estimation will be provided in \S4.4 where 
the spatial fluctuation is taken into account. 
 Note that the latter is reduced to eq.~\ref{eq:infinite-loop} 
in the limit $\xi_{\rm GL} \rightarrow \infty$.

\section{Microscopic Theory in the infinite-loop order}

 Hereafter, we investigate the role of vertex corrections 
on the basis of the repulsive Hubbard model. 
 The basic formulation has been provided in \S2 and \S3.

 Before going on to the main issue, we summarize the advantages of 
the microscopic theory. 
 (1) the properties of the SC fluctuation is microscopically derived.  
 This advantage has enabled us to discuss the doping dependence 
without using any phenomenological assumption. 
 (2) The renormalization of the quasi-particles is taken into account. 
 Owing to the electron correlation, the energy scale is renormalized 
to be smaller, and therefore the pseudogap occurs around 
$T \sim 0.01 t \sim 100{\rm K}$. 
 Note that the pseudogap in the attractive Hubbard model appears in 
much higher temperature region (see \S5.1). 
 (3) The $d$-wave symmetry of the superconductivity is 
naturally derived. 
 The characteristic momentum dependence of the superconducting gap 
is derived, which is different from the simple form 
$\Delta(k) \propto \cos k_{\rm x} - \cos k_{\rm y}$ especially in the 
electron-doped cuprates.~\cite{rf:yanasereview} 
 (4) The strong enhancement of the AF spin fluctuation is taken into 
account. The magnetic properties, which will be discussed in \S4.4 are 
dominated by the spin correlation around $\q \sim (\pi,\pi)$. 
 (5) The momentum dependence of quasi-particles is 
taken into account. 
 The momentum dependent life time plays an essential role 
for the understanding of transport phenomena 
in the pseudogap state~\cite{rf:yanaseTRPG}. 
 The remarkably short mean free path at ``hot spot'' is induced by 
the spin fluctuation, which provides a consistency with 
the assumption $\xi_{\rm GL} \gg l$.

 In the following, we focus on the higher order corrections 
with respect to the SC fluctuation. 
 The multiple scattering arising from the spin fluctuation has been 
investigated motivated by the indication of 
Schrieffer.~\cite{rf:schrieffer1995} 
 The previous studies~\cite{rf:monthoux1997,rf:chubukov1997} 
have shown that this correction enhances the effective vertex, 
however does not affect qualitative roles of the spin fluctuation. 
 We ignore the higher order coupling between the SC and 
spin fluctuations beyond the renormalization of Green function. 
 The interests are focused on the roles of SC fluctuation.

\subsection{Single-particle properties}

 In order to perform a infinite-loop calculation, we tentatively 
consider the neighborhood of \Tc and assume $\xi_{\rm GL} \gg l$. 
 Then, the Green function is expressed as eq.~\ref{eq:infinite-loop}.  
 Correspondingly, the self-energy in the $1$-loop and 
self-consistent $1$-loop approximation is expressed as, 
\begin{eqnarray}
\label{eq:FLEX+1-loop}
&& \hspace{-10mm} 
{\mit{\it \Sigma}}_{\rm S}(k) = \Delta(\k)^{2} G_{\rm F}(-k), 
\\
\label{eq:FLEX+sc-1-loop}
&& \hspace{-10mm} 
{\mit{\it \Sigma}}_{\rm S}(k) = \Delta(\k)^{2} G(-k), 
\end{eqnarray}
respectively.  
 We have ignored the frequency dependence of the coupling vertex 
$\phi(k)$, which gives no important effect in the vicinity of 
$T=T_{\rm c}$. 
 In this subsection, we ignore the feedback effects on the SC and spin 
fluctuations 
in order to fix fluctuation propagators in each approximation. 
 A fully self-consistent calculation including the feedback 
effect is performed in \S4.2-4. 
 We have confirmed that qualitatively same results are obtained in the 
self-consistent calculation.

\begin{figure}[htbp]
  \begin{center}
\includegraphics[height=10cm]{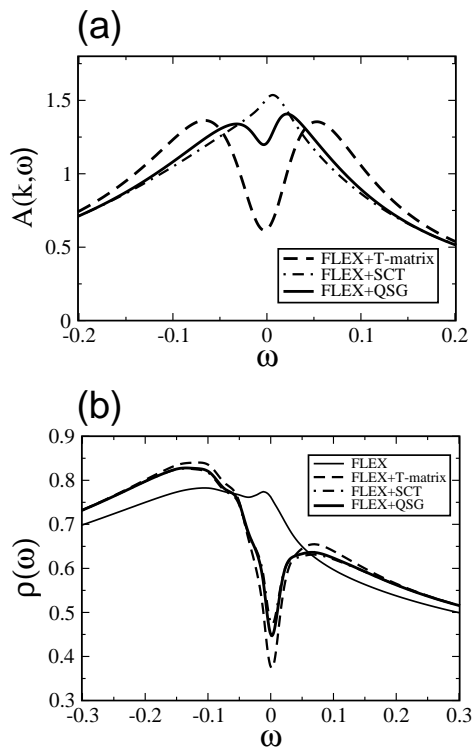}
    \caption{(a) The spectral function at the hot spot, 
                 $\k=(0.05 \pi,\pi)$.  
             (b) DOS. 
             The parameters are chosen as $T=T_{\rm c}^{\rm MF}$ 
             and $r=0.001$. 
             }
\label{fig:aboveTmf}
  \end{center}
\end{figure}

 Here, we take into account the inter-layer coupling in order to avoid
the singularity in the exactly two-dimensional system. The TDGL 
expansion for the Lawrence-Doniach model is adopted as, 
\begin{eqnarray}
  \label{eq:3DTDGL}
  t(\mbox{\boldmath$q$},0)=
  \frac{g}{t_{0} + b \mbox{\boldmath$q$}^{2} + 2 b r (1-\cos q_{\rm z})},  
\end{eqnarray}
 with introducing the phenomenological parameter $r$. 
 Here the TDGL parameters $t_{0}$ and $b$ are microscopically determined. 
 The phenomenological parameter is expressed by the anisotropy of the 
SC fluctuation as $r=\xi_{\rm z}^{2}/\xi_{0}^{2}$, where $\xi_{\rm z}$ is 
the coherence length along the {\it c}-axis in the unit of 
the inter-layer spacing. 
 The exactly two dimensional and isotropic three dimensional systems 
correspond to $r=0$ and $r=1$, respectively. 
 The small value of $r \ll 1$ is a underlying assumption of 
eq.~\ref{eq:3DTDGL}. 
 We choose $r=0.01 \sim 0.001$ as a typical value.

 We show the results of the single particle spectral function 
in Fig.~\ref{fig:aboveTmf}(a). 
 The expressions~\ref{eq:FLEX+1-loop}, \ref{eq:FLEX+sc-1-loop} and 
\ref{eq:infinite-loop} are denoted as FLEX+T-matrix, FLEX+SCT and 
FLEX+QSG, respectively. 
 It these estimations, the effect of SC fluctuation is 
represented by the single parameter $\Delta$. 
 We find that the temperature dependence of $\Delta$ is not significant 
around $T=T_{\rm c}$ owing to the inter-layer coupling. 
 Therefore, the single particle properties are almost temperature 
independent, while they are rapidly altered below 
$T_{\rm c}$.~\cite{rf:yanaseSC} 
 The experimental results have supported this 
result.~\cite{rf:renner,rf:miyakawa,rf:dipasupil,rf:ding,rf:shenPG,
rf:normanPG,rf:ARPESreview} 

 It is shown that the gap structure in the spectral function 
is suppressed in the FLEX+SCT, 
while it clearly appears in the non-self-consistent calculation 
(FLEX+T-matrix). 
 This is the discrepancy explained in \S2. 
 We see that the non-self-consistent $1$-loop approximation is 
qualitatively correct. 
 The calculation including the infinite order vertex corrections 
(FLEX+QSG) shows the pseudogap in the spectral function, 
although the gap structure is reduced from the lowest order approximation. 
 It is concluded that the self-consistent $1$-loop order theory 
under-estimates the effects of SC fluctuation, while the 
non-self-consistent one over-estimates them. 
 This qualitative conclusion will be clearly understood from 
the perturbative point of view (\S5.3).  
 It may be expected that the self-consistent $1$-loop approximation 
is always incorrect as is discussed for the single particle spectrum. 
 In the following, we show that this expectation is invalid and 
the SCFT is quite appropriate for the macroscopic quantities.

 In the present case, the pseudogap in the infinite-loop calculation 
is broadened by the competition with the Fermi liquid behavior 
arising from the spin fluctuation. 
 We note that the Fermi liquid behavior is basically obtained 
in the nearly AF Fermi liquid theory~\cite{rf:yanaseFLEXPG}, 
although some deviations from the conventional Fermi liquid are 
derived.~\cite{rf:moriyaAD,rf:yanasereview} 
 As a result, the difference between FLEX+SCT and FLEX+QSG is 
significantly reduced. This is an origin of the unimportance of 
vertex corrections for the macroscopic quantities.

 The DOS is a typical one, which is shown in Fig.~\ref{fig:aboveTmf}(b). 
 It is shown that the pseudogap is similarly obtained in three calculations. 
 We will show that the $1$-loop approximation provides 
a rather different result if we assume $s$-wave superconductivity 
and neglect the spin fluctuation (\S5.1). 
 In the present case, the difference between approximations is 
significantly reduced by the $d$-wave symmetry, which allows  
low-energy excitations, and by the spin fluctuation. 
 Thus, better convergence of the loop expansion is expected in this 
microscopic theory rather than the phenomenological models.

\subsection{Phase diagram}

 In this subsection, we estimate the doping dependence of \Tcc. 
 For this purpose we perform three kinds of calculation. 
 In the previous paper~\cite{rf:yanaseFLEXPG} 
we have performed the SCFT for the estimation of \Tc where 
eqs.~\ref{eq:anomalous-rpa-singlet}-\ref{eq:total-self-energy} are 
self-consistently solved. 
 Here we denote the calculation with eq.~\ref{eq:FLEX+sc-1-loop} 
instead of eq.~\ref{eq:FLEX+T-matrixap} as the same abbreviation. 
 In addition to that, we perform the self-consistent calculation with 
eq.~\ref{eq:infinite-loop} instead of eq.~\ref{eq:FLEX+T-matrixap} 
which is denoted as self-consistent FLEX+Gaussian approximation 
(SCFG). 
 The calculation with 
eq.~\ref{eq:FLEX+1-loop} instead of eq.~\ref{eq:FLEX+T-matrixap} 
is denoted as non-self-consistent FLEX+T-matrix approximation (NSCFT). 
 These calculations correspond to the self-consistent $1$-loop, 
infinite-loop and $1$-loop approximation, respectively.  
 In these calculations, the feedback effect on the spin fluctuation 
is self-consistently taken into account, which enhances the value of 
\Tcc.~\cite{rf:yanaseFLEXPG}

 In the previous paper~\cite{rf:yanaseFLEXPG}, the critical point 
has been determined by the criterion $\lambda_{\rm e}=1-\epsilon$ 
instead of $\lambda_{\rm e}=1$ as a phenomenological inclusion of  
the inter-layer coupling. 
 In the present study, we explicitly take into account the 
inter-layer coupling with use of eq.~\ref{eq:3DTDGL}. 
 The value of $r$ roughly corresponds to the phenomenological cut-off 
as $\epsilon \sim 4br$. Since the value of $b$ is typically $b \sim 5$, 
the previously used value $\epsilon=0.02$~\cite{rf:yanaseFLEXPG} 
corresponds to $r \sim 0.001$, which is relevant for high-\Tc cuprates. 
 Owing to the finite value of $r$, the logarithmic divergence in 
two-dimension is cut off and $\Delta$ at $T=T_{\rm c}$  
is obtained as 
\begin{eqnarray}
  \label{eq:3Ddelta}
\Delta^{2} = \frac{|g|T}{4 \pi b} {\rm Arccos} \frac{q_{\rm c}^{2}}{2 r}
\sim \frac{|g|T}{4 \pi b} \log \frac{\pi^{2}}{r}. 
\end{eqnarray}
 Here $q_{\rm c}$ is the cut off momentum along the $ab$-plane, which 
is chosen to be $q_{\rm c}=\pi$. 
 This procedure enables us to avoid the finite size effect 
which is serious around $T=T_{\rm c}$.

\begin{figure}[htbp]
  \begin{center}
\includegraphics[height=11cm]{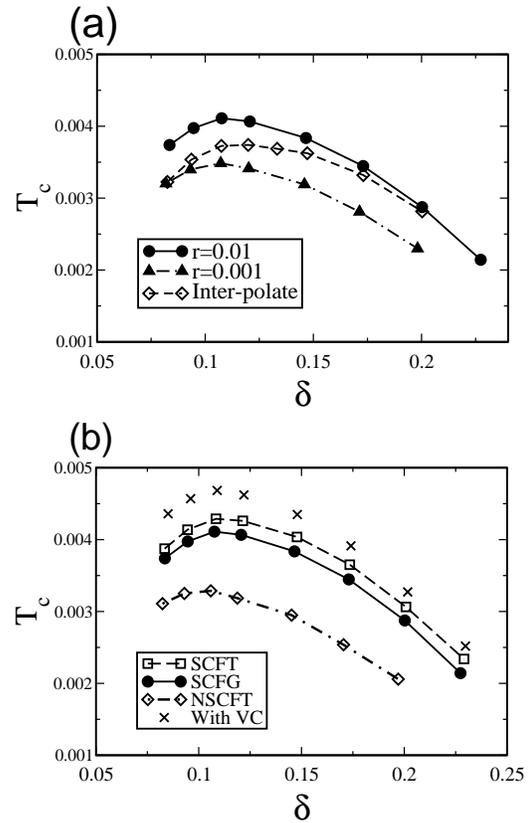}
    \caption{
             (b) Results of SCFG for various value of $r$. 
                 The diamonds are obtained by taking account of the 
                 doping dependence of inter-layer coupling. 
             (a) The doping dependence of \Tcc. 
                 The abbreviations, SCFT, SCFG, NSCFT have been 
                 explained in the text. 
                 The times shows the results including 
                 the vertex correction on the pairing correlation
                 function. The details and discussion will be given 
                 in \S4.3. 
             }
\label{fig:HubbardTc}
  \end{center}
\end{figure}

 First, we show the effect of inter-layer coupling 
in Fig.~\ref{fig:HubbardTc}(a) where the results of SCFG are shown. 
 Even if the value of $r$ differs by an order, the difference of \Tc 
does not exceed the factor $2$ in the whole doping range. 
 This is owing to the logarithmic dependence of $\Delta$ 
(see eq.~\ref{eq:3Ddelta}). 
 The qualitatively same doping dependence for $r=0.01$ and $r=0.001$ 
is an important property which allows a universal nature of 
the pseudogap~\cite{rf:idouniversal} independent of the inter-layer structure. 
 Although the value of $r$ is remarkably small 
in BSCCO,~\cite{rf:iyereview,rf:basovrelation} 
this difference affects only quantitatively.

 It is shown that \Tc has a maximum around $\delta \sim 0.11$, 
while the critical temperature in the mean field theory \Tmf 
develops with decreasing $\delta$ in the under-doped region 
(see Ref.36). 
 This qualitatively different behavior between \Tc and \Tmf is 
consistent with our understanding of the phase 
diagram.~\cite{rf:yanasereview} 
 That is, the onset temperature of the pseudogap follows \Tmf 
below which the SC fluctuation becomes active more and more. 
 On the contrary, the \Tc is suppressed in the under-doped region 
owing to the fluctuation. 
 We will show that the energy scale of the pseudogap is 
generally determined by \Tmf and not by \Tcc, independent of 
the inter-layer coupling (\S5.1).

 It is predominantly believed that the inter-layer coupling is reduced 
with decreasing the doping. 
 We have shown in Fig.~\ref{fig:HubbardTc}(a) the result including the 
variation of $r$. 
 Then, we have chosen the linear interpolation 
$r=0.001+0.06 (\delta-0.08)$ so as to be $r=0.001$ at $\delta=0.08$ 
and $r=0.01$ at $\delta=0.23$.  
 We see that the value of optimal doping is increased 
by the variation of inter-layer coupling.

 We see from Fig.~\ref{fig:HubbardTc}(b) that the higher order 
corrections do not change the qualitative behavior of \Tcc. 
 Three approximations give similar results. 
 In particular, the close results are obtained between SCFT and SCFG. 
 Therefore, it is concluded that vertex corrections beyond SCFT 
has no important role for the estimation of \Tcc. 
 This is not owing to the details of microscopic theory, since quite 
the same conclusion is obtained for the attractive Hubbard model 
(\S5.1). 
 Note that the feedback effect through the spin fluctuation 
furthermore reduces the differences. 
 Since the pseudogap in the single particle excitations is 
under-estimated in the SCFT, the \Tc is higher than SCFG. 
 However, it will be shown that the \Tc in the SCFT is 
still under-estimated compared with the calculation including the 
vertex correction on the pairing correlation function. 
 Details are explained in the next subsection.

\subsection{Vertex correction to the pairing correlation function}

 Thus far, we have investigated the vertex corrections 
on the Green function, which are shown 
in Fig.~\ref{fig:yokohigh}. 
 Actually, more careful discussion is necessary in calculating 
the many body correlation function. 
 The Maki-Thompson (MT) term and Aslamasov-Larkin (AL) term on the 
electric conductivity are typical ones which have been investigated 
since 1970's. 
 These corrections on the magnetic properties have been investigated 
in Refs.82-84 where the unimportance of MT-term in the $d$-wave case 
has been pointed out.   
 The microscopic treatment on the electric conductivity and 
Hall coefficient has been performed on the basis of the SCFT 
approximation.~\cite{rf:yanaseTRPG} 
 Then, it has been shown that the dominant correction is the 
indirect feedback effect because of the characteristic momentum 
dependence of high-\Tc cuprates. Then, the unimportance of the MT and 
AL terms has been concluded in contrast to the knowledges 
for the weak coupling superconductors. 
 The correction to the Nernst effect has been also investigated 
on the basis of the SCFT approximation~\cite{rf:kontaniPG} where the 
MT term plays an important role.

\begin{figure}[htbp]
  \begin{center}
\includegraphics[height=3.7cm]{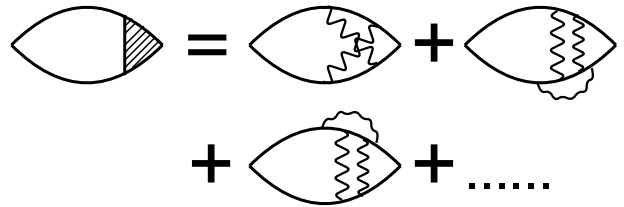}
    \caption{The diagrammatic representation for the vertex corrections 
             on the irreducible pairing susceptibility. 
             }
\label{fig:vertexontwo}
  \end{center}
\end{figure}

 Here, we investigate the vertex corrections to the pairing correlation 
function which is equivalent to the T-matrix except for a finite 
constant. 
 The pairing correlation function determines the SC fluctuation, 
and furthermore the value of \Tcc. 
 Since we find that the role of SC fluctuation is robust as is 
expected in \S3, we focus on the correction to \Tcc. 
 We investigate corrections to the irreducible susceptibility in the 
Cooper channel, which are shown in Fig.~\ref{fig:vertexontwo}. 
 These corrections are regarded as a contraction of the 
mode coupling higher than the sixth order.  
 However, it is easily confirmed that these corrections include the 
lowest order term in the sense of loop expansion, rather than 
the terms studied previously~\cite{rf:tokumitu,rf:sato}.

 In order to estimate the corrections in Fig.~\ref{fig:vertexontwo} 
as well as those in Fig.~\ref{fig:yokohigh}, 
we collect the all order terms by using the same approximation as in 
\S4.1 and \S4.2. Then, the $q$-dependence of Green functions is ignored. 
 This simplification enables us to obtain a closed expression, which 
is the \eli equation modified as, 
\begin{eqnarray}
  \label{eq:modifiedeliashberg}
   \lambda_{{\rm e}} \phi(k) =
  -\sum_{k'} V_{{\rm a}}(k,k') P(k') \phi(k'). 
\end{eqnarray}
 Correspondingly, the function $\lambda(q)$ is obtained 
in the following way, 
\begin{eqnarray}
   \lambda(q)=
  -\sum_{k,p}\phi^{*}(k) V_{\rm a}(k-p) Q(p,q) \phi(p). 
  \label{eq:modifiedT-matrix}
\end{eqnarray}
 Here, the functions $P(k)$ and $Q(p,q)$ are defined as, 
\begin{eqnarray}
&&  \hspace{-15mm} 
P(k)=\int_{0}^{\infty}
\frac{t e^{-t}}{|G_{\rm F}(k)|^{-2} + t \Delta(\k)^{2}} {\rm d}t, 
\label{eq:irrducibleT-matrixzero}
\\
&&  \hspace{-15mm} Q(p,q)=\frac{G_{\rm F}(p) G_{\rm F}(q-p)}
{\Delta(\p)^{2} |G_{\rm F}(p)|^{2} - \Delta(\p-\q)^{2} |G_{\rm F}(q-p)|^{2}} 
\nonumber
\\
&& \hspace{0mm} 
\times
\int_{0}^{\infty} \{ \frac{\Delta(\p)^{2}}
{|G_{\rm F}(p)|^{-2} + t \Delta(\p)^{2}} 
\nonumber \\
&& \hspace{5mm}
-\frac{\Delta(\p-\q)^{2}}{|G_{\rm F}(q-p)|^{-2} + t \Delta(\p-\q)^{2}} \}
e^{-t} {\rm d}t, 
\label{eq:irrducibleT-matrix}
\end{eqnarray}
respectively.

 The result of \Tc has been shown in Fig.~\ref{fig:HubbardTc}(b). 
 We understand that the \Tc in the present calculation is larger than 
that obtained in the SCFT. 
 Thus, the vertex corrections represented by Fig.~\ref{fig:vertexontwo} 
are larger than those included in the renormalization of Green 
function (Fig.~\ref{fig:yokohigh}).

 Note that this conclusion is expected from the perturbative 
point of view. 
 We see that the lowest order vertex correction included in  
Fig.~\ref{fig:yokohigh} is in the $3$-loop order. 
 On the other hand, the $2$-loop order term exists in the corrections 
represented by Fig.~\ref{fig:vertexontwo}. 
 It is understood from the simple estimation that this $2$-loop order 
term enhances \Tcc. 
 Therefore, the result in this subsection is consistent with the 
perturbation theory where the lowest order term is believed to determine 
the qualitative tendency.

 We have confirmed that the TDGL parameter $b$ increases owing to the 
correction in Fig.~\ref{fig:vertexontwo}. 
 Therefore, the SC fluctuation is reduced by these corrections, while 
the increase of \Tc enhances the thermal fluctuation. 
 We have confirmed that this effect is not serious, quantitatively. 
 Therefore, we ignore the corrections in Fig.~\ref{fig:vertexontwo} 
in the following discussion about the magnetic properties. 
 We close this subsection by noting that the qualitative feature 
of the doping dependence is still not altered, as shown in 
Fig.~\ref{fig:HubbardTc}(b).

\subsection{Magnetic properties}

 At the last of this section, we show the temperature dependence of 
NMR $1/T_{1}T$ and uniform susceptibility $\chi_{\rm s}(0)$. 
 They are typical quantities showing the pseudogap.~\cite{rf:yasuoka,
rf:warren,rf:takigawa1991,rf:itoh1992,rf:itoh1998,rf:ishida1998,rf:alloul} 
 In order to discuss the magnetic properties, we perform another 
calculation proposed by Sadovskii~\cite{rf:sadovskii}.  
 While we have only to consider the situation $\xi_{\rm GL} \gg l$ 
for the estimation of \Tcc, the spatial fluctuation may be an essential 
nature above \Tcc. 
 Sadovskii's method has provided a simple and reasonable estimation 
for the spatial fluctuation. 
 A justification will be obtained by the comparison with the numerical 
estimation within the $4$-loop order. 
 A Monte Carlo simulation is a method of use, as has been 
performed for the phenomenological model with spin 
fluctuation and that with charge fluctuation.~\cite{rf:monthouxPG} 
 However, we carry out an approximate but analytic treatment 
in order to obtain a clearer understanding and to compare several 
approximations in an equal footing.

 In general, a spatial fluctuation gives rise to the broadening of 
self-energy in each order. 
 In the method proposed by Sadovskii, this effect is taken into account 
by linearizing the inverse of Green function as 
$G_{\rm F}(k-q) = 1/({\rm i}\omega_{n}-\varepsilon(\k)
+\tilde{\vv}(\k)\q - \Sigma_{\rm F}(k))$ and replacing the 
$\q$-summation with adding an imaginary quantity. 
 The imaginary quantity is chosen to be the inverse of 
typical length scale, namely $\xi_{\rm GL}^{-1}$. 
 Then, the self-energy in the $n$-loop order is expressed as, 
\begin{eqnarray}
  \label{eq:sadovskii}
&& \hspace{-10mm}
\Sigma^{(n)}(k)=|\Delta(\k)|^{2 n} \Pi_{i=1}^{2 n + 1} 
 ({\rm i}\omega_{n} - (-1)^{i} \varepsilon(\k) 
\nonumber \\
&& \hspace{0mm}
- (-1)^{i} \Sigma_{\rm F}((-1)^{i} k) 
\pm  {\rm i} p(i) \tilde{v}(\k) \xi_{\rm GL}^{-1})^{-1}. 
\end{eqnarray}
 The integer $p(i)$ is the number of SC fluctuation propagators 
enclosing the Green function appearing in the $i$-th order. 
 The sign $+$ ($-$) should be chosen in the upper (lower) half plane. 
 The quasi-particle velocity is obtained as 
$\tilde{v}(\k)=|\tilde{\vv}(\k)|$ where  
$\tilde{\vv}(\k)=\vv(\k) + 
\partial{\rm Re}\Sigma_{\rm F}(\k,{\rm i} \pi T)/\partial \k$.

 This method was first proposed for one-dimensional 
charge fluctuation systems as an exact solution~\cite{rf:sadovskii}. 
 Although this procedure has been proved to be not exact 
beyond $1$-loop order,~\cite{rf:tchernyshyov1D} a good accuracy as an 
approximation has been numerically confirmed in case of the 
complex order parameter.~\cite{rf:kopietzPG,rf:millisPG} 
 Here this method is applied to the two-dimensional SC fluctuation. 

 The infinite summation of eq.~\ref{eq:sadovskii} is expressed 
as a recursive fraction,  
\begin{eqnarray}
  \label{eq:recursivefraction}
&& \hspace{-10mm}
\Sigma_{\rm S}(k) = 
\\
&& \hspace{-10mm}
\frac{\Delta(\k)^{2}}
  {\displaystyle -G_{\rm F}^{-1}(-k) +{\rm i} n(1) 
                  \tilde{v}(\k) \xi_{\rm GL}^{-1} + 
             \frac{\Delta(\k)^{2}}
  {\displaystyle G_{\rm F}^{-1}(k) +{\rm i} n(2) 
                  \tilde{v}(\k) \xi_{\rm GL}^{-1} 
  + .....
             }}, 
\nonumber
\end{eqnarray}
where $n(i)=[(i+1)/2]$ and $[...]$ is the Gauss symbol. 
 It is easily confirmed that this expression is attributed to 
eq.~\ref{eq:infinite-loop} in the limit 
$\xi_{\rm GL} \rightarrow \infty$.  
 On the other hand, this expression is attributed to the zeroth order theory 
in the opposite limit $\xi_{\rm GL} \rightarrow 0$. 
 Thus, the expression \ref{eq:recursivefraction} can be regarded as an 
interpolation between the limiting cases.

 It should be noted that the applicability of this method to the 
two dimensional case will need more discussion. 
 In the previous application for AF spin
fluctuation~\cite{rf:schmalianPG} and CDW fluctuation~\cite{rf:kuchinskii}, 
the displacement of the fluctuation propagator 
$\frac{1}{\xi_{\rm GL}^{-2}+\q^{2}} \rightarrow 
\frac{\xi_{\rm GL}^{-1}}{\xi_{\rm GL}^{-2}+\q_{\parallel}^{2}} 
\frac{\xi_{\rm GL}^{-1}}{\xi_{\rm GL}^{-2}+\q_{\perp}^{2}} $ 
is implicitly performed. 
 This displacement is accurate in the interested region 
$q \sim 1/\xi_{\rm GL} $. 
 Owing to this procedure, the momentum summation is attributed to the 
one-dimensional problem. 
 Therefore, the accuracy of Sadovskii's method is similarly expected 
in the two-dimensional case. 
 It should be stressed that the one-dimension is not particular  
for the phenomenological Gaussian fluctuation model, while 
the quantum fluctuation is essential in the one-dimensional 
microscopic models. 
 Note that two or higher dimensional models where the Sadovskii's method 
is exact have been proposed~\cite{rf:fujimotoPG,rf:kuchinskii}.

\begin{figure}[htbp]
  \begin{center}
\includegraphics[height=7cm]{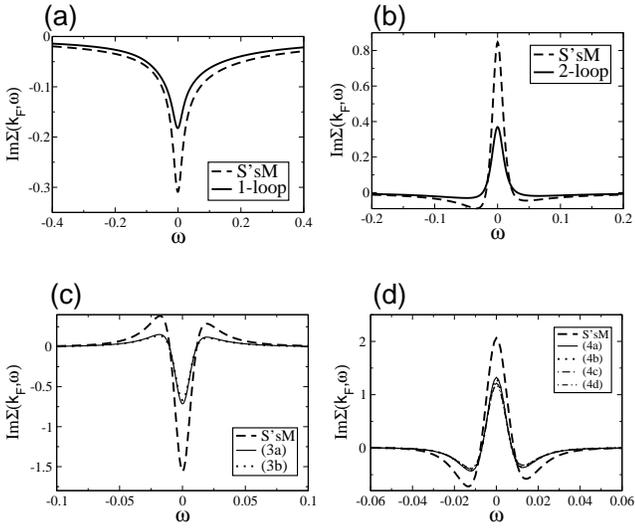}
    \caption{Results of the loop expansion. 
             (a), (b), (c) and (d) show the imaginary part of 
             the self-energy in the $1$-, $2$-, $3$- and $4$-loop order,
             respectively. 
             The dashed lines denoted as S'sM show the results of 
             Sadovskii's method. 
             The indices (3a), (3b) and (4a-4d) correspond to 
             the diagrams in Fig.~\ref{fig:yokohigh}. 
             The parameters are chosen as $U/t=3.2$ and $T=0.0084$ 
             where $T=1.035 T_{\rm c}^{\rm MF}$.             
             }
\label{fig:loop-e}
  \end{center}
\end{figure}

 In order to provide another ground, we perform the loop expansion 
up to the $4$-loop order. 
 Then, the quasi-static approximation is adopted and the dominator of 
Green function is linearized. 
 The imaginary part of the self-energy in each order is estimated 
by the numerical integration. 
 The result of the Sadovskii's method (eq.~\ref{eq:sadovskii}) is shown 
together in Fig.~\ref{fig:loop-e}. 
 We see the qualitatively same behaviors between the Sadovskii's method 
and naive loop-expansion in each order.

 Another interesting finding is that the crossing diagram and 
non-crossing one are almost equivalent in the results of loop expansion. 
 For example, the difference between Fig.~\ref{fig:yokohigh}(3a) and 
Fig.~\ref{fig:yokohigh}(3b) 
is almost invisible in Fig.~\ref{fig:loop-e}(c). 
 We see the same feature in the $4$-loop order terms. 
 This feature is obtained even if we do not linearize the dominator of 
Green function.~\cite{rf:yanasepro} 
 It should be stressed that this feature is an essential assumption of 
the Sadovskii's method. 
 The results shown in Fig.~\ref{fig:loop-e} have confirmed that 
this assumption is satisfied with remarkable accuracy. 
 From these results we believe that Sadovskii's method is appropriate 
at least for qualitative discussions, although quantitative accuracy 
has not been proved.

 The numerical factor $n(i)=[(i+1)/2]$ in eq.~\ref{eq:recursivefraction} 
is precisely obtained by counting the diagrams. 
 This factor is replaced with $n(i)=1$ in the self-consistent $1$-loop 
approximation (SCFT) and with $n(i)=\delta_{i,1}$ in the $1$-loop 
approximation (NSCFT). 
 The role of vertex corrections are clarified by the comparison between 
these calculations.

\begin{figure}[htbp]
  \begin{center}
\includegraphics[height=11cm]{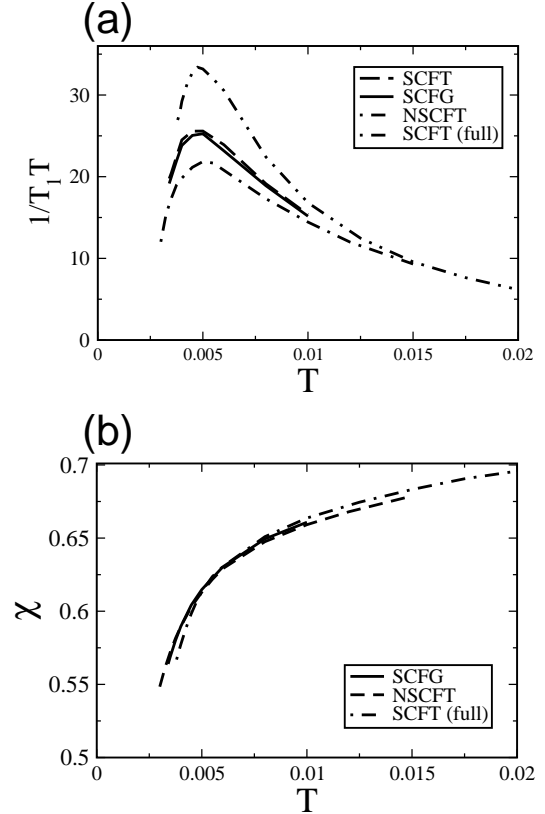}
    \caption{(a) The results of NMR $1/T_{1}T$ obtained 
                 by several approximations. 
             (b) The uniform susceptibility. 
             We denote the SCFT approximation fully including the
             spatial and dynamical fluctuations as SCFT (full). 
             }
\label{fig:magnetic}
  \end{center}
\end{figure}

 We show the results of NMR $1/T_{1}T$ and uniform spin 
susceptibility in Fig.~\ref{fig:magnetic}. 
 Here the estimation of $\sum_{q<q_{\rm c}}|t(q,0)|$ is 
numerically performed with the cut-off $q_{\rm c}=\pi/2$. 
 Because the inter-layer coupling is not important except for the 
vicinity \Tcc, we perform the calculation in the two-dimension.  
 The NMR $1/T_{1}T$ is estimated by using the fluctuation-dissipation 
theorem, which is expressed as 
\begin{eqnarray}
 \label{eq:NMR}
 1/T_{1}T =  \sum_{\mbox{\boldmath$q$}} F_{\perp}(\mbox{\boldmath$q$})
             [\frac{1}{\omega}
             {\rm Im} \chi_{{\rm s}}^{{\rm R}} (\mbox{\boldmath$q$}, \omega)
             \mid_{\omega \to 0}],
\end{eqnarray}
 where 
\begin{eqnarray}
F_{\perp}(\mbox{\boldmath$q$}) = \frac{1}{2}\{A_{1} + 2 B
(\cos q_{x}+\cos q_{y})\}^{2} + 
\nonumber \\
\frac{1}{2}
\{A_{2} + 2 B (\cos q_{x}+\cos q_{y})\}^{2}.  
\end{eqnarray}
 The hyperfine coupling constants $A_{1}, A_{2}$ and $B$ are chosen as
$A_{1} = 0.84 B$ and $A_{2} = -4 B$ \cite{rf:barzykin}.
 We have added the results obtained by the SCFT without using the 
quasi-static approximation. 
 Then, the momentum and frequency summations are performed 
by using the fast Fourier transformation (FFT) and taking account of 
the same cut-off $q_{\rm c}=\pi/2$.  

 It is clearly shown that the pseudogap phenomena appear in all of the 
calculations in a similar way. 
 NMR $1/T_{1}T$ shows a maximum above \Tcc. 
 On the other hand, the uniform susceptibility decreases with 
decreasing temperature from much higher temperature. 
 This qualitatively different behavior is one of the characteristics of 
under-doped high-\Tc cuprates.~\cite{rf:timusk,rf:yasuoka,rf:odaspin} 
 This is owing to the interplay of the AF spin fluctuation and 
SC fluctuation.

 We see that the vertex correction beyond SCFT is almost negligible. 
 The unimportance of the vertex correction in the present case 
is very significant compared with the previous subsections. 
 The result for $1/T_{1}T$ is particularly surprising because 
this quantity is very sensitive to the approximation, owing to 
the strong enhancement of the spin fluctuation. 
 This fact indicates an extremely precise cancellation of the 
vertex corrections for magnetic properties.

 It is noted that the same conclusion is obtained by neglecting the 
$q$-dependence of the Green function as is performed in \S4.1-3.   
 Then, the effect of spatial fluctuation is obviously neglected. 
 This nature of approximation should be contrasted with 
the Sadovskii's method where the spatial fluctuation is 
over-estimated.~\cite{rf:tchernyshyov} 
 It is expected that the unimportance of vertex corrections 
for the magnetic properties is a robust conclusion, 
since it is confirmed by two complementary approximations.

 At the last of this section, we comment on the closing of the 
pseudogap. As has been shown in \S4.1, the pseudogap appears in the 
spectral function around $T=T_{\rm c}$. 
 As the temperature increases, the gap structure is gradually 
broadened. In the present parameters, the pseudogap at the hot spot 
disappears around $T=0.005$ which corresponds to the maximum of NMR 
$1/T_{1}T$. We have confirmed that the closing of the pseudogap is 
yielded by the spatial fluctuation but not by the decrease of $\Delta$. 
 Indeed, the temperature dependence of $\Delta$ is weak since 
the increase of temperature enhances the thermal fluctuation. 
 Therefore, the closing of the pseudogap occurs as a blurring but not 
as a shrinking. 
 This is quite consistent with the experimental observation in ARPES. 
 Note that the quasi-static approximation is not valid far above \Tcc. 
 Then, the dynamical fluctuation is also important as well as 
the spatial fluctuation.

\section{Phenomenological Theory: More General Aspects}

 In this section, the comparison to spin fluctuation and charge 
fluctuation is performed on the basis of the phenomenological theory. 
 We will see that the qualitative role of vertex corrections discussed 
in \S4 is non-trivial and determined by the symmetry of order parameter. 
 An analysis of the infinite order calculation 
within the phenomenological theory has been reported for the 
charge fluctuation~\cite{rf:sadovskii}, 
AF spin fluctuation~\cite{rf:schmalianPG} and  
SC fluctuation~\cite{rf:tchernyshyov,rf:fujimotoPG}, respectively. 
 Here, we provide a comprehensive view and shed light on the essential 
character of SC fluctuation.

\subsection{In case of SC fluctuation}

 First, we summarize the phenomenological theory for the SC fluctuation. 
 The results on the single particle properties are clearly understood. 
 In this section, we ignore the $k$-dependence of $\Delta(k)$ and 
phenomenologically assume the value of $\Delta$ for simplicity. 
 This situation corresponds to the attractive Hubbard model which is 
briefly discussed at the last of this subsection. 
 Since another scattering process is not taken into account here, 
the retarded Green function in the $1$-loop, self-consistent $1$-loop 
and infinite-loop calculations are described, respectively, 
\begin{eqnarray}
  \label{eq:p1-loop}
&& \hspace{-14mm}
  G_{1}^{{\rm R}}(\mbox{\boldmath$k$},\omega)=\frac{\omega+\varepsilon(\k)}
  {(\omega+\varepsilon(\k))
  (\omega-\varepsilon(\k))-\Delta^{2}}
\\
  \label{eq:psc1-loop}
&& \hspace{-14mm}
  G_{1{\rm s}}^{{\rm R}}(\mbox{\boldmath$k$},\omega)=
  \frac{1}{\omega-\varepsilon(\k)}
  \frac{2}{1+
  \sqrt{1-\frac{4 \Delta^{2}}{\omega^{2}-\varepsilon(\k)^{2}}}}
\\
  \label{eq:pinfinite-loop}
&&  \hspace{-14mm}
  G_{\infty}^{{\rm R}}(\mbox{\boldmath$k$},\omega) =
  \int_{0}^{\infty} \frac{\omega+\varepsilon(\k)}
  {(\omega+\varepsilon(\k))(\omega-\varepsilon(\k))-t \Delta^{2}} 
  \hspace{2mm} e^{-t} {\rm d}t.
\\
&&
\hspace{-12mm}
 = \int_{0}^{\infty} 
  \frac{\omega+\varepsilon(\k)}
  {(\omega+\varepsilon(\k))(\omega-\varepsilon(\k))-x^{2}} 
  \hspace{2mm} e^{-x^{2}/\Delta^{2}} \frac{2 \pi x}{\pi \Delta^{2}} {\rm d}x.  
\end{eqnarray}
 Strictly speaking, the condition $l \ll \xi_{\rm GL}$ is not relevant 
in the present case since $l=\infty$. 
 However, these expressions are sufficient for the qualitative 
discussion.

 The Green functions in eqs.~\ref{eq:p1-loop} and \ref{eq:pinfinite-loop} 
imply the relation to the normal Green function in the superconducting 
state. 
 We see that eq.~\ref{eq:p1-loop} is equivalent to the 
normal Green function in the BCS theory. 
 The square of the order parameter in the latter is replaced 
by the thermal weight of the fluctuation. 
 The expression in the infinite-loop calculation corresponds to 
the normal Green function averaged by the Gaussian distribution 
of the order parameter. 
 Quasi-particles propagate under the fluctuating order parameter 
which is approximated to be classical and Gaussian in the present case. 
 If the mean free path is shorter than the correlation length, 
the spatial fluctuation of the order parameter is not effective for 
quasi-particles. 
 Therefore, a physically relevant result is obtained 
as eq.~\ref{eq:pinfinite-loop}. 
 Thus, the quasi-static approximation provides a clear interpretation. 
 On the other hand, it is difficult to find a clear interpretation of 
self-consistent $1$-loop approximation (see eq.~\ref{eq:psc1-loop}).  
 This fact implies a potentiality that higher order terms are 
inappropriately taken into account in this approximation.

\begin{figure}[htbp]
  \begin{center}
\includegraphics[height=10cm]{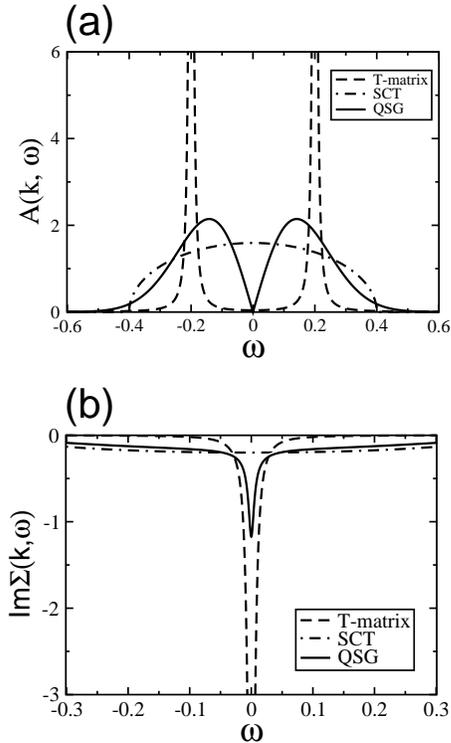}
    \caption{(a) The spectral function on the Fermi surface 
                 ($\varepsilon(\k)=0$) obtained by 
                 eqs.~\ref{eq:p1-loop}-\ref{eq:pinfinite-loop}. 
             (b) The imaginary part of the self-energy. 
                 We have chosen the parameter as $\Delta=0.2$. 
             }
\label{fig:3-calc}
  \end{center}
\end{figure}

 This possibility has been realized in \S4.1 where 
the single particle spectral function has been discussed. 
 We illustrate the spectral function obtained from 
eqs.~\ref{eq:p1-loop}-\ref{eq:pinfinite-loop} (Fig.~\ref{fig:3-calc}(a)). 
 The failure of the self-consistent $1$-loop order theory is clearly 
understood. 
 The single peak structure appears in the self-consistent 
$1$-loop approximation (SCT), while the double peak structure is 
clearly observed in the $1$-loop (T-matrix) and infinite-loop (QSG) 
approximations.~\cite{rf:tchernyshyov} 
 The gap structure in the QSG is broad compared with the $1$-loop 
approximation owing to the Gaussian distribution of the pair field. 
 Thus, the result obtained in \S4.1 is qualitatively understood from 
this phenomenological theory.

 This observation implies that the vertex corrections significantly 
cancel the higher order corrections included in the SCT.  
 Note that the SCT applied to the attractive Hubbard model is 
one of the conserving approximations formulated by Baym and Kadanoff. 
 The present calculation is an example that the preservation of some 
conservation low does not necessarily improve the accuracy of 
the calculation. 
 The TPSC approximation has been proposed as an improvement where 
some conservation laws unsatisfied in the SCT are satisfied 
phenomenologically.~\cite{rf:kyungd-wave,rf:kyungTPSC} 
 Then, the property of fluctuations and the coupling vertex to 
them are basically discussed. 
 However, the present diagrammatic method indicates that the 
improvement of SCT should be performed by including the multiple 
scattering like Fig.~\ref{fig:yokohigh}.

 The characteristics of each approximation will be clarified by showing 
the self-energy in Fig.~\ref{fig:3-calc}(b). 
 It is shown that the anomalous scattering peak around $\omega=0$ 
almost disappears in the SCT. 
 This is the reason of the broad single peak in the spectral function. 
 The scattering peak remains in the infinite order calculation,  
while it is reduced from the $1$-loop order theory.

\begin{figure}[htbp]
  \begin{center}
\vspace{2mm}
\includegraphics[height=4.5cm]{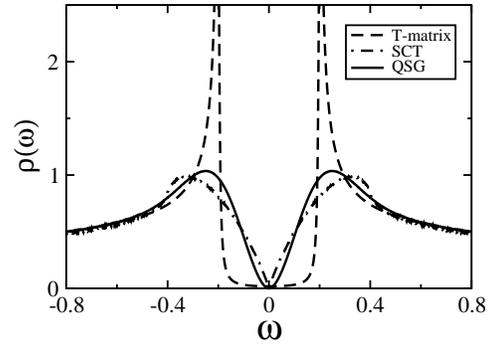}
    \caption{Results of the DOS. The dispersion relation 
             eq.~\ref{eq:high-tc-dispersion} is used with $t'/t=0$ and 
             $n=1$. 
             }
\label{fig:3-calc2}
  \end{center}
\end{figure}

 Here, we show the DOS in Fig.~\ref{fig:3-calc2}. 
 We see that the SCT is appropriate with regard to the DOS. 
 This is because the spectral function slightly apart from the Fermi 
surface is not so different between the SCT and QSG. 
 On the other hand, the result of the T-matrix approximation is 
rather different from that of QSG. 
 This discrepancy has not been shown in the microscopic theory 
(Fig.~\ref{fig:aboveTmf}(b)). 
 Then, the $1$-loop order theory has been much improved 
owing to the $d$-wave symmetry and to the scattering from spin 
fluctuation.

\begin{figure}[htbp]
  \begin{center}
\includegraphics[height=10cm]{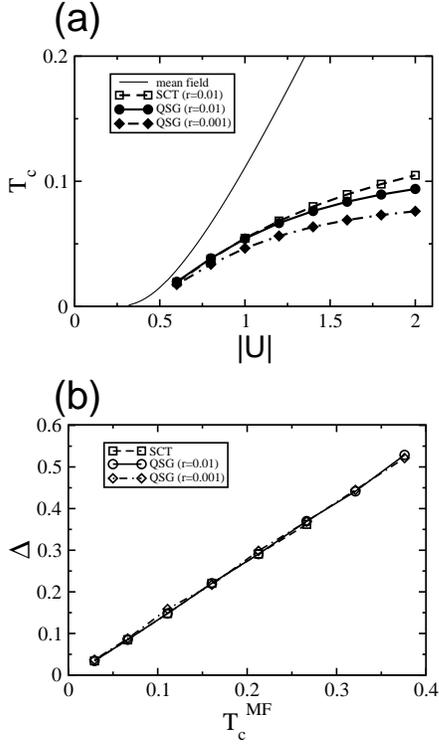}
    \caption{(a) \Tc in the attractive Hubbard model. 
                 We show the results of SCT at $r=0.01$ and QSG 
                 at $r=0.01$ and $r=0.001$, respectively. 
                 The thin solid line show \Tc in the mean field theory. 
             (b) The relation between \Tmf and $\Delta$ at $T=T_{\rm c}$ 
                 in each calculation and parameter. 
             }
\label{fig:attractive}
  \end{center}
\end{figure}

 At the last of this subsection, we provide results for the attractive 
Hubbard model in Fig.~\ref{fig:attractive}. 
 Then, the Green function is expressed as 
eqs.~\ref{eq:p1-loop}-\ref{eq:pinfinite-loop} and the value of $\Delta$ 
and the propagator of fluctuation are self-consistently determined. 
 Here the inter-layer coupling is taken into account with use of 
eq.~\ref{eq:3DTDGL}. 
 Fig.~\ref{fig:attractive}(a) shows that the vertex 
correction beyond the SCT is not serious for the estimation of \Tcc. 
 This conclusion is in common to the microscopic theory in \S4.2.

 The relation between the excitation gap 
$ \Delta $ at $T=T_{\rm c}$ and 
the transition temperature in the mean field theory \Tmf is 
shown in Fig.~\ref{fig:attractive}(b). 
 We clearly see the universal behavior independent of the 
indicator of three-dimensionality $r=(\xi_{\rm z}/\xi_{0})^{2}$. 
 \Tmf and $\Delta$ has a universal relation, 
although the value of \Tc depends on the inter-layer coupling. 
 This is an important finding for the whole understanding of 
cuprate superconductors. 
 The result in Fig.~\ref{fig:attractive}(b) shows that the energy scale 
of the pseudogap is not determined by \Tcc, but by \Tmf, which is robust 
for the variation of the inter-layer structure.

 It should be noted that this relation is derived in the renormalized 
Gaussian fluctuation region focused in this paper. 
 Therefore, the experimental observation is not an evidence 
for the fixed amplitude expected in the phase fluctuation 
region.~\cite{rf:emery,rf:kwon,rf:franz,rf:phase,rf:eckl,rf:tesanovic}

\subsection{Comparison to spin and charge fluctuations}

 Next, the possibility of the pseudogap 
phenomena induced by spin fluctuation or charge fluctuation 
is discussed in order to extract the essential properties of SC fluctuation. 
 The possibility on the AF spin fluctuation has been investigated 
extensively in the studies of high-\Tc 
cuprates.~\cite{rf:kampfPG,rf:chubukovPG,rf:vilkAFPG,rf:schmalianPG} 
 Although we will stress the invalidity of the quasi-static 
approximation for the spin fluctuation (Appendix A), this approximation 
is used for a comparison. 
 The role of charge fluctuation has been interested in the context 
of the stripe order.~\cite{rf:tranquada} 
 In our opinion, the charge fluctuation is not important 
for a unified understanding of high-\Tc superconductors.  
 However, the charge fluctuation is an important example in the 
present purpose because the role of vertex corrections is 
qualitatively different. 
 The difference between the charge and spin fluctuations 
has been discussed by the Monte Carlo simulation.~\cite{rf:monthouxPG}  
 We show that the SC fluctuation is classified into the same class 
as that of the spin fluctuation.

 Here, the discussion is restricted to the commensurate case 
where the ordering vector is $(\pi,\pi)$. 
 The infinite series for the Green function is obtained 
in the same manner as in \S3. 
 The results are expressed as, 
\begin{eqnarray}
  \label{eq:infinite-series-C}
&& \hspace{-10mm} 
G_{\rm c}(k) = \sum_{n} \frac{(2 n - 1)!!}{2^{n}} \Delta_{\rm c}^{2n}  
             G^{(0)}(k)^{n+1} G^{(0)}(k+Q)^{n} \hspace{5mm} 
\nonumber \\
&& \hspace{35mm}
       ({\rm charge  \hspace{2mm} fluctuation}),
\\
  \label{eq:infinite-series-S}
&& \hspace{-10mm} 
 G_{\rm s}(k) = \sum_{n}  \frac{(2 n + 1)!!}{2^{n}} \Delta_{\rm s}^{2n}
             G^{(0)}(k)^{n+1} G^{(0)}(k+Q)^{n} \hspace{5mm} 
\nonumber \\
&& \hspace{38mm}
       ({\rm spin  \hspace{2mm} fluctuation}), 
\end{eqnarray}
respectively. 
 The estimation of the numerical factor is not easy in case of 
the spin fluctuation since not only the longitudinal mode but also the 
transverse one exist. 
 We obtain the factor $\frac{(2 n + 1)!!}{2^{n}}$ 
from the recurrence formula. 
 Here we have defined the thermal weight of the charge and 
spin fluctuation as 
$\Delta_{\rm c}^{2}= g_{\rm c}^{2} T \sum_{q} \chi_{\rm c}(Q+q,0)$ and 
$\Delta_{\rm s}^{2}= g_{\rm s}^{2} T \sum_{q} \chi_{\rm s}(Q+q,0)$ 
where $g_{\rm c}$ and $g_{\rm s}$ are the effective coupling constants. 
 In the RPA or FLEX for the Hubbard model, 
these coupling constants are equivalent to $U$. 
 Summing up the infinite series, the Green function is obtained as, 
\begin{eqnarray}
    \label{eq:pinfinite-loop-C}
&& \hspace{-10mm}
G_{\rm c}^{{\rm R}}(\mbox{\boldmath$k$},\omega) =
 \sqrt{\frac{2}{\pi}}  \int_{0}^{\infty}
  \frac{\omega-\varepsilon(\k+\Q)}
  {(\omega-\varepsilon(\k))(\omega-\varepsilon(\k+\Q))-
  t^{2} \Delta_{\rm c}^{2}/2} 
  \hspace{2mm} 
\nonumber \\
&& \hspace{20mm}
\times e^{-t^{2}/2} {\rm d}t,  
\\
&& 
\hspace{5mm}
= \int_{0}^{\infty} 
  \frac{\omega-\varepsilon(\k+\Q)}
  {(\omega-\varepsilon(\k))(\omega-\varepsilon(\k+\Q))-x^{2}} 
  \hspace{2mm} 
\nonumber \\
&& \hspace{20mm}
\times 
e^{-x^{2}/\Delta_{\rm c}^{2}} 
  \frac{2}{\sqrt{\pi \Delta_{\rm c}^{2}}}{\rm d}x,  
\\
    \label{eq:pinfinite-loop-S}
&& \hspace{-10mm} 
 G_{\rm s}^{{\rm R}}(\mbox{\boldmath$k$},\omega) =
 \sqrt{\frac{2}{\pi}}  \int_{0}^{\infty} 
  \frac{\omega-\varepsilon(\k+\Q)}
  {(\omega-\varepsilon(\k))(\omega-\varepsilon(\k+\Q))-
  t^{2} \Delta_{\rm s}^{2}/2} 
  \hspace{2mm} 
\nonumber \\
&& \hspace{20mm}
\times
t^{2} e^{-t^{2}/2} {\rm d}t, 
\\
&& \hspace{5mm}
= \int_{0}^{\infty} 
  \frac{\omega-\varepsilon(\k+\Q)}
  {(\omega-\varepsilon(\k))(\omega-\varepsilon(\k+\Q))-x^{2}} 
  \hspace{2mm} 
\nonumber \\
&& \hspace{20mm}
\times 
  e^{-x^{2}/\Delta_{\rm s}^{2}} 
  \frac{4 \pi x^{2}}{(\pi \Delta_{\rm s}^{2})^{3/2}}{\rm d}x.  
\end{eqnarray}
 These expressions are interpreted as the Green function in the ordered 
state averaged by the Gaussian distribution.~\cite{rf:schmalianPG}  
 The essential difference in the formula is the power of $x$ in the 
integrand, which is determined by the symmetry of the order parameter. 
 The charge fluctuation has the $Z_{2}$-symmetry, while the SC 
fluctuation and spin fluctuation has the $U(1)$- and $SU(2)$-symmetry, 
respectively. 
 They are translated into the Ising, XY and Heisenberg symmetry, 
respectively.

\begin{figure}[htbp]
  \begin{center}
\includegraphics[height=7cm]{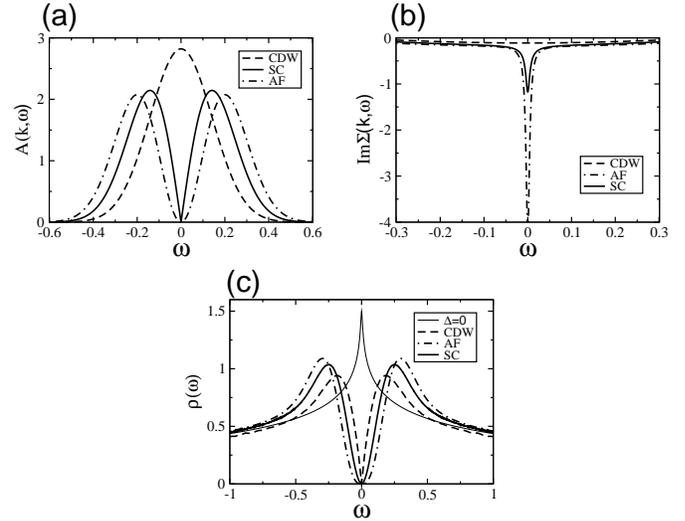}
    \caption{(a) The spectral function on the Fermi surface 
                 obtained by the infinite-loop 
                 calculation for charge, SC and spin fluctuations, 
                 respectively. 
             (b) The imaginary part of the self-energy. 
             (c) The DOS. 
                 The parameters are chosen as 
                 $\Delta=\Delta_{\rm c}=\Delta_{\rm s}=0.2$, 
                 $t/t=0$ and $n=1$. 
                 We have shown the DOS in the non-interacting case for 
                 a comparison. 
             }
\label{fig:3-type}
  \end{center}
\end{figure}

 The spectral function obtained from eqs.~\ref{eq:pinfinite-loop-C} 
and ~\ref{eq:pinfinite-loop-S} are shown in Fig.~\ref{fig:3-type}(a). 
 We see that the pseudogap is not observed in case of the charge 
fluctuation, while it is observed in case of the spin fluctuation. 
 This is easily understand from the fact that the function 
$\exp(-x^{2})$ has the maximum at $x=0$, while the functions 
$x \exp(-x^{2})$ and  $x^{2} \exp(-x^{2})$ have the maximum at $|x|>0$. 
 This is the reason why the SC fluctuation and spin fluctuation are 
classified into the same class. 
 Fig.~\ref{fig:3-type}(b) shows that the scattering peak leading 
to the pseudogap in the spectral function almost disappears 
in case of the charge fluctuation.

 It should be stressed that qualitatively the same result is obtained 
for three kinds of fluctuations, when we adopt the $1$-loop or 
self-consistent $1$-loop approximation. 
 The differences obtained above are essentially owing to the role of  
vertex corrections.  
 It is understood from the above discussion that the qualitative role 
of vertex corrections essentially depends on the symmetry of 
order parameter. 
 In this sense, the results on the SC fluctuation are non-trivial. 
 The $U(1)$-symmetry of the order parameter plays an essential role. 
 In case of the charge fluctuation, the appropriate result 
is obtained from the self-consistent $1$-loop approximation rather than 
the non-self-consistent one.

 It is shown in Fig.~\ref{fig:3-type}(c) that the charge fluctuation 
induces the gap structure of DOS, which has been focused in the previous 
studies~\cite{rf:leericeanderson,rf:sadovskii,rf:millisPG,rf:kopietzPG,
rf:monienPG,rf:kuchinskii}. 
 Thus, the pseudogap in the quantities including the momentum summation 
is robust for the properties of order parameter as well as 
those of approximations. 
 The suppression of DOS around $\omega=0$ is 
more remarkable when the symmetry of the order parameter is higher. 

 Before closing this section, we comment on the realistic problem 
for high-\Tc cuprates. Then, the pseudogap induced by the AF spin 
fluctuation is not clearly observed in the FLEX approximation. 
 We consider that this is not owing to the nature of the FLEX 
approximation, which is the self-consistent $1$-loop order theory, 
but owing to the dynamical fluctuation which is important as 
will be shown in Appendix A.

\subsection{Perspective from the perturbation theory}

\begin{figure}[htbp]
  \begin{center}
\includegraphics[height=9cm]{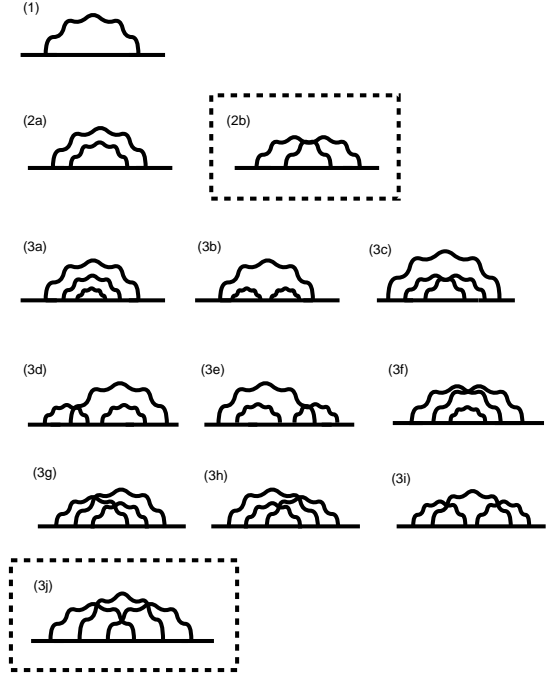}
    \caption{The diagrammatic representation of the self-energy within 
             the $3$-loop order. 
             The term (2b) is the lowest order vertex correction in 
             case of the charge or spin fluctuation. 
             The term (3j) is the lowest order vertex correction in 
             case of the SC fluctuation. 
             }
\label{fig:vertexcorrection}
  \end{center}
\end{figure}

 The qualitatively different role of the vertex corrections discussed 
in \S5.2 is clarified from the perturbative point of view. 
 Fig.~\ref{fig:vertexcorrection} shows the diagrammatic representation of 
the loop expansion in case of the spin or charge fluctuation. 
 We have shown the diagrams within the $3$-loop order. 
 Note that Figs.~\ref{fig:vertexcorrection}(1), (2a), (3a) and (3b) 
are included in the self-consistent $1$-loop approximation, 
which is the FLEX approximation~\cite{rf:FLEX} in the microscopic 
theory. 
 The other terms are classified into the vertex correction.

 In order to make a discussion clear, we introduce a common expression 
for each term. 
 In case of the charge and spin fluctuation, we denote   
\begin{eqnarray}
  \label{eq:definition-cs}
&& \hspace{-10mm}
\Sigma^{(2{\rm a})}(k)=\sum_{q_{1},q_{2}} V(q_{1}) V(q_{2}) 
\nonumber \\ 
&& \hspace{5mm}
\times
G(k+q_{1}) G(k+q_{1}+q_{2}) G(k+q_{1}),
\\ 
&& \hspace{-10mm} 
\Sigma^{(2{\rm b})}(k)=\sum_{q_{1},q_{2}} V(q_{1}) V(q_{2}) 
\nonumber \\ 
&& \hspace{5mm}
\times
G(k+q_{1}) G(k+q_{1}+q_{2}) G(k+q_{2}), 
\\
&& \hspace{-10mm}
\Sigma^{(3{\rm a})}(k)=\sum_{q_{1},q_{2},q_{3}} V(q_{1}) V(q_{2}) V(q_{3}) 
G(k+q_{1})
\nonumber \\ 
&& \hspace{5mm}
\times
G(k+q_{1}+q_{2}) G(k+q_{1}+q_{2}+q_{3}) 
\nonumber \\ 
&& \hspace{5mm}
\times
G(k+q_{1}+q_{2}) G(k+q_{1}), 
\\
&& \hspace{-10mm}
\Sigma^{(3{\rm j})}(k)=\sum_{q_{1},q_{2},q_{3}} V(q_{1}) V(q_{2}) V(q_{3}) 
G(k+q_{1})
\nonumber \\ 
&& \hspace{5mm}
\times
G(k+q_{1}+q_{2}) G(k+q_{1}+q_{2}+q_{3}) 
\nonumber \\ 
&& \hspace{5mm}
\times
G(k+q_{2}+q_{3}) G(k+q_{3}), 
\end{eqnarray}
and so on. 
 The vertex $V(q)$ is defined as 
$V(q)=\frac{1}{2} g_{\rm c}^{2} \chi_{\rm c}(q)$ 
and $V(q)=\frac{3}{2} \ g_{\rm s}^{2} \chi_{\rm s}(q)$ for 
charge and spin fluctuation, respectively. 
 We have chosen the notation in which the coefficient of the 
terms included in the self-consistent $1$-loop approximation is unity. 
 It is easily confirmed that odd order terms generally introduce 
the anomalous contribution leading to the pseudogap, 
while even order terms reduce it.

 We see that the lowest order vertex correction is in the second 
order (Fig.~\ref{fig:vertexcorrection}(2b)). 
 In case of the charge fluctuation, all of the terms contribute to the 
self-energy with a coefficient $1$. 
 Then, the self-energy in the second order is 
$\Sigma^{(2)}(k)=\Sigma^{(2{\rm a})}(k)+\Sigma^{(2{\rm b})}(k)$ 
and that in the third order is 
$\Sigma^{(3)}(k)=\Sigma_{\alpha={\rm a}}^{{\rm j}} 
\Sigma^{(3{\rm \alpha})}(k)$. 
 In this case, the lowest order vertex correction reduces the anomalous 
contribution. 
 Therefore, the effect of the charge fluctuation is still overestimated 
in the self-consistent $1$-loop approximation.

 On the other hand, the coefficients are complicated 
in case of the spin fluctuation.  
 We obtain the second order term as 
$\Sigma^{(2)}(k)=\Sigma^{(2{\rm a})}(k)-\frac{1}{3} 
\Sigma^{(2{\rm b})}(k)$. 
 The negative coefficient $-1/3$ arises from the coupling between the 
longitudinal and transverse modes. 
 Thus, the sign of the lowest order vertex correction is opposite to the 
case of charge fluctuation. 
 Therefore, the qualitatively different role of the vertex correction is 
expected. 
 The self-consistent $1$-loop approximation underestimates the 
effects of spin fluctuation. 
 Note that the $1$-loop approximation overestimates them 
since the vertex correction term $-\frac{1}{3} \Sigma^{(2{\rm b})}(k)$ 
has smaller coefficient $1/3$ than $\Sigma^{(2{\rm a})}(k)$.  
 The Monte Carlo study has reported the qualitatively consistent result 
about the difference between the charge and spin 
fluctuations~\cite{rf:monthouxPG}.

 The numerical factor $-1/3$ has played an important role 
in the study of the pairing 
interaction.~\cite{rf:monthoux1997,rf:chubukov1997,rf:caprara} 
 The lowest order vertex correction enhances the pairing interaction 
mediated by the spin fluctuation,~\cite{rf:monthoux1997,rf:chubukov1997} 
while it reduces the pairing interaction mediated by 
the charge fluctuation~\cite{rf:caprara}. 
 Thus, the spin fluctuation is likely to be effective for the 
anisotropic superconductivity rather than the charge fluctuation, 
owing to the vertex corrections as well as the numerical factor $3$.

 At last, we clarify the case of SC fluctuation. 
 Then, the diagrams in Figs.~\ref{fig:vertexcorrection}(2b) and (3c-i) 
are absent. 
 The number of the vertex correction terms is remarkably small 
in this case. 
 Indeed, this is a characteristic property of the complex order 
parameter. 
 We introduce the expressions for the remaining terms as, 
\begin{eqnarray}
  \label{eq:definition-sc}
&& \hspace{-5mm} 
\Sigma^{(2{\rm a})}(k)=\sum_{q_{1},q_{2}} V(q_{1}) V(q_{2}) 
\nonumber \\
&& \hspace{5mm}
\times
G(-k+q_{1}) G(k-q_{1}+q_{2}) G(-k+q_{1}),
\\
  \label{eq:definition-sc3a}
&& \hspace{-5mm} 
\Sigma^{(3{\rm a})}(k)=\sum_{q_{1},q_{2},q_{3}} V(q_{1}) V(q_{2}) V(q_{3}) 
\nonumber \\
&& \hspace{5mm}
\times
G(-k+q_{1}) G(k-q_{1}+q_{2}) G(-k+q_{1}-q_{2}+q_{3}) 
\nonumber \\
&& \hspace{5mm}
\times G(k-q_{1}+q_{2}) G(-k+q_{1}), 
\\
  \label{eq:definition-sc3j}
&& \hspace{-5mm} 
\Sigma^{(3{\rm j})}(k)=\sum_{q_{1},q_{2},q_{3}} V(q_{1}) V(q_{2}) V(q_{3}) 
\nonumber \\
&& \hspace{5mm}
\times
G(-k+q_{1}) G(k-q_{1}+q_{2}) G(-k+q_{1}-q_{2}+q_{3}) 
\nonumber \\
&& \hspace{5mm}
\times
G(k+q_{2}-q_{3}) G(-k+q_{3}),
\end{eqnarray}
where $V(q)=t(q,0)$. 
 The second and third order terms are obtained as 
$\Sigma^{(2)}(k)=\Sigma^{(2{\rm a})}(k)$ and 
$\Sigma^{(3)}(k)=\Sigma^{(3{\rm a})}(k)+\Sigma^{(3{\rm b})}(k)+
\Sigma^{(3{\rm j})}(k)$, respectively. 
 Because the coefficient of each term is unity, 
the lowest order vertex correction $\Sigma^{(3{\rm j})}(k)$ has 
qualitatively the same role as $\Sigma^{(3{\rm a})}(k)$ which enhances 
the pseudogap phenomena. 
 Thus, the role of vertex corrections clarified in \S4  
is expected from the perturbative point of view. 
 Because the vertex correction in this case is higher order than 
that of spin and charge fluctuations, it is expected to be less important. 
 This is an underlying origin of the unimportance of the vertex
correction on the macroscopic quantities.

\begin{table}[htbp]
  \begin{center}
    \begin{tabular}{|c||c|c|c|} \hline 
 & charge & SC & spin \\\hline
2nd order & $1$ & $0$ & $-1/3$ \\\hline
3nd order & $8$ & $1$ & $-22/9$ \\\hline
    \end{tabular}
    \caption{The coefficient of vertex correction terms.
             The cases of the charge, SC and spin fluctuation are shown, 
             respectively. 
             We show the coefficient of the second order term 
             $\Sigma^{(2{\rm b})}$ 
             and the summation of the coefficients of third order 
             terms $\Sigma^{(3{\rm c})}$-$\Sigma^{(3{\rm j})}$. 
             }
  \end{center}
\end{table}

 Summarizing, the results obtained in \S5.1 and \S5.2 are qualitatively 
consistent with the expectation from the perturbation theory. 
 The qualitatively different role of vertex corrections is explained 
from the lowest order vertex correction, even if the contribution from 
the higher order term is larger. 
 Thus, the perturbation theory provides another understandings for the 
results, although we have to perform an infinite order calculation 
in order to obtain the results in a closed form.

\section{Summary and Discussion}

 In this paper, we have investigated the role of higher order 
corrections beyond the T-matrix approximation for SC fluctuation. 
 The results obtained from the detailed analysis are summarized in the 
following way. 
 The first one is the importance of the vertex correction for the 
single particle spectral function. 
 Then, the self-consistent T-matrix approximation is qualitatively 
incorrect, while the non-self-consistent one is appropriate. 
 The second one is the unimportance of the vertex correction for the 
macroscopic quantities. 
 We have explicitly estimated the DOS, superconducting transition 
temperature, NMR $1/T_{1}T$ and magnetic susceptibility on the 
basis of the repulsive Hubbard model. 
 It has been shown that the self-consistent $1$-loop order theory 
is rather precise for these quantities. 
 We have found that the cancellation of the vertex correction 
for the magnetic properties is especially precise. 
 These results basically provide a justification of the $1$-loop 
order theory, although a care is necessary for the interpretation of 
ARPES measurements.  
 The present study provides a clear point of view to the previous 
studies within the $1$-loop order calculation.~\cite{rf:janko,rf:yanasePG,
rf:metzner,rf:perali3D,rf:yanasereview,rf:yanaseFLEXPG,rf:yanaseTRPG}

 We have pointed out that the $d$-wave symmetry of the superconductivity 
and the renormalization from the spin fluctuation significantly reduce 
the effect of vertex corrections. 
 This fact is natural since the higher order terms are 
generally suppressed by these properties. 
 Therefore, a better convergence of the loop expansion is expected 
in the realistic situation rather than in the situation adopted in 
the phenomenological models.

 It should be stressed again that the justification of the $1$-loop 
order theory is obtained even when the naive perturbation calculation 
fails. 
 Then, the higher order corrections are renormalized by summing up 
infinite series. 
 Note that the cancellation occurs between the even order terms and 
odd order ones which have qualitatively opposite behaviors at low energy. 
 The same order terms enhance rather than cancel each other. 
 Therefore, the calculation within the finite order is 
remarkably inaccurate. 
 The results in $3$-loop order, $4$-loop order..... seriously 
oscillate. 
 The $1$-loop order theory is significantly precise rather than the 
other finite order calculation. 
 However, we have pointed out in \S5.3 that the qualitative roles of 
vertex corrections are generally determined by the lowest order term. 
 The unimportance of the vertex corrections arising from the 
SC fluctuation rather than that from the charge and spin fluctuations
is also expected from the perturbative point of view. 
 Thus, the perturbation theory is still valid in this qualitative sense.

 Note that the $1$-loop order theory has been used widely in the 
fluctuation theory including the FLEX 
approximation.~\cite{rf:FLEX,rf:moriyaAD,rf:yanasereview} 
 The simplicity of the $1$-loop order theory 
has enabled us to investigate rich issues in a coherent way. 
 At the moment, there is no conclusive evidence for the validity 
because so-called Migdal theorem is not generally applicable. 
 It may be considered that this problem is serious 
in the strongly correlated electron systems 
because the strong coupling nature of the fluctuation is a 
characteristic property of them. 
 However, the present study implies a wide applicability of the 
$1$-loop order theory, although it inevitably fails in the deeply  
critical region. 
 For example, we believe that the $1$-loop order theory for 
the spin fluctuation is more effective than expected from 
the lowest order estimation for the vertex 
correction.~\cite{rf:monthoux1997,rf:chubukov1997,rf:schmalianPG} 
 In many cases, the contribution which is not included in the 
fluctuation theory is more important rather than the vertex correction,  
as a quantitative correction.~\cite{rf:yanasereview} 
 We expect that the observation in the present study will be a typical 
example in the rich field of the fluctuation theory in the 
correlated electron systems.

\section*{Acknowledgements}

 The authors are grateful to Professors M. Ogata and K. Yamada 
for fruitful discussions. 
 Numerical computation in this work was partly carried out 
at the Yukawa Institute Computer Facility. 
 The present work was supported by a Grant-In-Aid for Scientific 
Research from the Ministry of Education, Science, Sports and Culture, Japan.

\appendix

\section{On the validity of quasi-static approximation}

\begin{figure}[htbp]
  \begin{center}
\includegraphics[height=5cm]{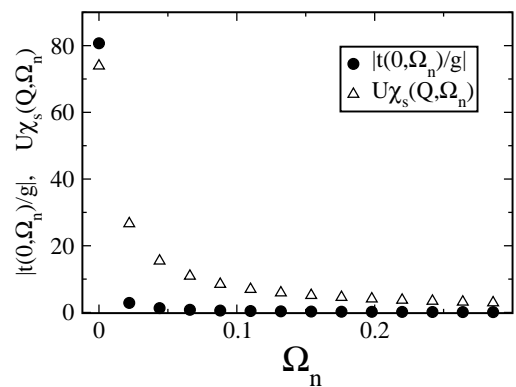}
    \caption{The Matsubara frequency dependence of the T-matrix 
             at $\q=0$ and that of the spin susceptibility 
             at $\q=(\pi,\pi)$. 
             The functions $|t(q)/g|$ and $U \chi_{\rm s}(q)$
             are shown for an equitable comparison. 
             The SCFG is adopted with $T=0.0035$. 
             }
\label{fig:quasi-static}
  \end{center}
\end{figure}

 This Appendix is provided as a ground for the quasi-static 
approximation which is used in this paper. 
 For a quantitative estimation needed for this purpose, 
we show the results obtained in the microscopic theory developed 
in \S4.4. 
 Then, the quasi-static approximation has been used in combination with 
Sadovskii's method. We show the result of SCFG at $T=0.0035$. 
 Fig.~\ref{fig:quasi-static} shows the $\Omega_{n}$-dependence of the 
pairing correlation function at $\q=0$. 
 The spin correlation function at $\q=(\pi,\pi)$ is shown for a comparison. 
 In order to perform a comparison in an equal footing, 
we show the functions $|t(q)/g|=|\lambda(q)/(1-\lambda(q))|$ and 
$U \chi_{\rm s}(q) = U \chi_{0}(q)/(1-U \chi_{0}(q)) $ and choose the 
parameters so that the values at $\Omega_{n}=0$ are almost equivalent.

 It is clearly shown that the pairing correlation function rapidly 
decreases for a finite Matsubara frequency. 
 Therefore, it is expected that the dynamical part of the T-matrix  
does not play any important role. 
 This conclusion has been confirmed also in the explicit calculation 
within the $1$-loop order theory (\S4.4). 
 Then, the SCFT approximations with and without quasi-static 
approximation provide similar results for magnetic properties. 
 we have confirmed this fact also for the single particle spectral 
function, although the result has not been shown.

 On the other hand, the spin correlation function takes not so small 
value at finite Matsubara frequency. 
 In other words, the characteristic frequency of the spin fluctuation 
remains to be larger than the temperature. 
 Thus, it is concluded that the quasi-static approximation for the 
spin fluctuation is not valid for the relevant region 
in the pseudogap state. 
 We wish to stress again that such a remarkable difference between the SC 
fluctuation and spin fluctuation is not owing to the enhancement factors 
$1/(1-\lambda(q))$ and $1/(1-U \chi_{0}(q)) $, but owing to the 
essential property of each fluctuation. 
 As has been explained in \S3.2, the characteristic frequency of the 
fluctuation is expressed as $\omega_{\rm sc} \sim T \varepsilon$ 
and $\omega_{\rm s} \sim E_{\rm F} \varepsilon_{\rm s}$ 
for SC fluctuation and for spin fluctuation, respectively. 
 We see a smaller coefficient in case of the SC fluctuation. 
 Indeed, this is one of the characteristics of the phase transition 
with logarithmic divergence at $T \rightarrow 0$. For example, the 
superconductivity and SDW with perfect nesting are the cases.

 It should be noted that the validity of the quasi-static approximation 
is related to the formation of the pseudogap in a straightforward manner. 
 While the anomalous contribution leading to the pseudogap arises from 
the quasi-static part, the dynamical part basically smears it. 
 The validity of the quasi-static approximation is actually 
an underlying origin of the clear pseudogap induced by the SC 
fluctuation.

 Note that the invalidity of the quasi-static approximation 
for the spin fluctuation is concluded in the pseudogap state, 
but it is not a general consequence. 
 This approximation for the spin fluctuation theory will be appropriate 
in higher temperature region and/or sufficiently close 
to the magnetic instability. 
 In the present case, the pseudogap in the magnetic excitation itself 
is an origin of the invalidity, since it means the reduced damping 
of spin fluctuation.

\section{Self-energy represented by Fig.~\ref{fig:highervertex}(f)}

\begin{figure}[htbp]
  \begin{center}
\includegraphics[height=5cm]{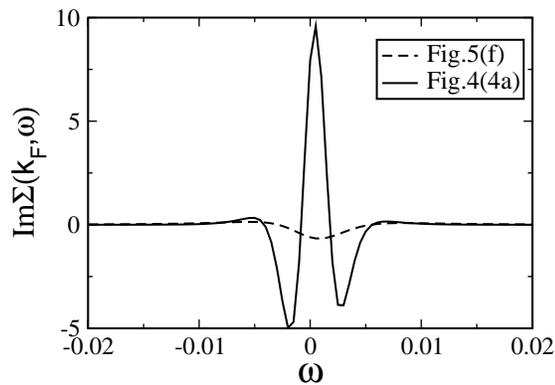}
    \caption{The imaginary part of the self-energy. 
             The dashed line shows the contribution from the
             non-Gaussian term represented by 
             Fig.~\ref{fig:highervertex}(f). 
             The solid line shows the self-energy represented by 
             Fig.~\ref{fig:yokohigh}(4a). 
             The temperature is chosen as $T=0.004$. 
             }
\label{fig:non-Gaussian}
  \end{center}
\end{figure}

 In this paper, we have ignored the contribution from the non-Gaussian 
terms which include higher order mode couplings. 
 Although considerable part of them is represented by the 
renormalization of fluctuation propagator 
(see  Figs.~\ref{fig:highervertex}(b) and (c)),  
the role of remaining terms is not clear. 
 The lowest order term included in the remaining part 
has been shown in Fig.~\ref{fig:highervertex}(f).
 Here, we provide a ground for our disregard of these terms. 
 
 We explicitly estimate the self-energy Fig.~\ref{fig:highervertex}(f) 
and compare with the $4$-loop order term in Fig.~\ref{fig:yokohigh}. 
 Here, we determine the T-matrix and Green function by using the SCFG 
as is explained in \S4.4. 
 For the estimation of Fig.~\ref{fig:highervertex}(f), we simply 
ignore the frequency and momentum dependences of the $4$th order vertex 
$\Gamma$ represented by Fig.~\ref{fig:highervertex}(d), 
as is done in the Ginzburg-Landau theory. 
 Since there are two equivalent diagrams of 
Fig.~\ref{fig:highervertex}(f) owing to the spin index, 
we multiplies the non-Gaussian terms by the factor $2$.

 The result is shown in Fig.~\ref{fig:non-Gaussian}. 
 It is understood that the non-Gaussian term is much smaller than the 
same order vertex corrections taken into account in this paper. 
 Because there are much more terms in the $4$-loop order in 
Fig.~\ref{fig:yokohigh} and they are almost equivalent (see \S4.4), 
the unimportance of the non-Gaussian term is remarkable in the $4$-loop 
order. 
 This is partly owing to the fact that the $4$-th order vertex 
$\Gamma=\Sigma_{k} |G(k)|^{4}$ is significantly reduced by 
the self-energy arising from the spin fluctuation as well as the SC 
fluctuation. 
 Although the role of higher order non-Gaussian terms is still 
unclear, we believe that they do not seriously affect the 
qualitative conclusions obtained in this paper.


\begin{thebibliography}{9}
\bibitem{rf:commentattractive}
Results on the attractive model with $d$-wave symmetry has been 
summarized in \S4.2 of Ref.36.
\end{thebibliography}
\end{document}